\documentclass[english,onecolumn,draftcls]{IEEEtran}
\usepackage[T1]{fontenc}
\usepackage[latin9]{inputenc}
\usepackage{verbatim}
\usepackage{amsthm}
\usepackage{amsmath}
\usepackage{amssymb}
\usepackage{graphicx}
\usepackage{esint}

\makeatletter
\theoremstyle{plain}
\newtheorem{thm}{\protect\theoremname}
\theoremstyle{plain}
\newtheorem{lem}[thm]{\protect\lemmaname}
\theoremstyle{plain}
\newtheorem{cor}[thm]{\protect\corollaryname}
\theoremstyle{remark}
\newtheorem{rem}[thm]{\protect\remarkname}

\makeatother

\usepackage{babel}
\providecommand{\corollaryname}{Corollary}
\providecommand{\lemmaname}{Lemma}
\providecommand{\remarkname}{Remark}
\providecommand{\theoremname}{Theorem}

\begin{document}

\title{Verification \hspace{-1mm}Decoding \hspace{-1mm}of \hspace{-1mm}High-Rate
\hspace{-1mm}LDPC \hspace{-1mm}Codes with Applications in Compressed
Sensing\\
{\normalsize (CLN 9-610 First Revision 01/22/12)} }

\author{Fan Zhang and Henry D. Pfister\\
{\normalsize Department of Electrical and Computer Engineering, Texas
A\&M University}\\
{\normalsize \{fanzhang,hpfister\}@tamu.edu}}
\maketitle
\begin{abstract}
This paper considers the performance of $(j,k)$-regular low-density
parity-check (LDPC) codes with message-passing (MP) decoding algorithms
in the high-rate regime. In particular, we derive the high-rate scaling
law for MP decoding of LDPC codes on the binary erasure channel (BEC)
and the $q$-ary symmetric channel ($q$-SC). For the BEC and a fixed
$j$, the density evolution (DE) threshold of iterative decoding scales
like $\Theta(k^{-1})$ and the critical stopping ratio scales like
$\Theta(k^{-j/(j-2)})$. For the $q$-SC and a fixed $j$, the DE
threshold of verification decoding depends on the details of the decoder
and scales like $\Theta(k^{-1})$ for one decoder.

Using the fact that coding over large finite alphabets is very similar
to coding over the real numbers, the analysis of verification decoding
is also extended to the the compressed sensing (CS) of strictly-sparse
signals. A DE based approach is used to analyze the CS systems with
randomized-reconstruction guarantees. This leads to the result that
strictly-sparse signals can be reconstructed efficiently with high-probability
using a constant oversampling ratio (i.e., when the number of measurements
scales linearly with the sparsity of the signal). A stopping-set based
approach is also used to get stronger (e.g., uniform-in-probability)
reconstruction guarantees. \end{abstract}

\section{Introduction\label{sec:Introduction}}

Compressed sensing (CS) is a relatively new area of signal processing
that has recently received a large amount of attention. The main idea
is that many real-world signals (e.g., those sparse in some transform
domain) can be reconstructed from a relatively small number of linear
 measurements. Its roots lie in the areas of statistics and signal
processing \cite{Chen-siamscicomp98,Donoho-it06,Candes-it06}, but
it is also very much related to previous work in computer science
\cite{Cormode-ja05,Gilbert-stoc07} and applied mathematics \cite{cdd-igpm06,JL-cm84,Gluskin-mus84}.
CS is also very closely related to error correcting codes, and can
be seen as source coding using linear codes over real numbers \cite{Sarvotham-isit06,Sarvotham-tr06,Baron-sp10,Xu-itw07,ZP-ita08,Zhang-aller08,dai-it-lmp}. 

In this paper, we analyze the performance of low-density parity-check
(LDPC) codes with verification decoding \cite{Luby-it05} as applied
to CS. The resulting approach is almost identical to that of Sudocodes
\cite{Sarvotham-isit06}, but our new perspective allows one to numerically
compute sparsity thresholds for a broad class of measurement matrices
under verification-based decoding. Changing the ensemble of measurement
matrices also allows an unbounded reduction in the oversampling ratio
relative to Sudocodes. A scaling approach is adopted to derive simple
expressions for the sparsity threshold as it approaches zero. Since
many interesting applications of CS involve very sparse (or compressible)
signals, this is a very interesting regime. From a coding perspective,
this corresponds to the high-rate limit and our results also have
implications for verification-based decoding of LDPC codes over large
finite fields. 

The analysis of CS in this paper is based on the noiseless measurement
of strictly-sparse signals \cite{cdd-igpm06,Candes-it06,Sarvotham-isit06}.
In the real world, the measurement process may introduce noise and
reconstruction algorithms must be implemented with finite-precision
arithmetic. Although the verification decoder discussed in this paper
is unstable in the presence of noise, this does not imply that its
performance analysis is not useful. The verification decoder can be
seen as a suboptimal version of list-message passing decoder \cite{zp-it-lmp},
which itself can be seen as a high-SNR limit of the full belief-propagation
(BP) decoder for CS \cite{Sarvotham-tr06,Baron-sp10}. Ideally, one
would study the BP decoder directly, but the DE analysis technique
remains intractable for decoders that pass functions as messages.
Still, we expect that a successful analysis of the BP decoder would
show that its performance is lower bounded by the verification decoder.

{} 

Sparse measurement matrices and message-passing reconstruction algorithms
for CS were introduced in \cite{Sarvotham-isit06,Sarvotham-tr06}.
Both ideas have since been considered by a number of other authors
\cite{Xu-itw07,ZP-ita08,BI-rpt08,BGIK-aller08,LMP-aller08,Donoho-pnas09,DMM-itw101,DMM-itw102,GI-procieee10}.
For example, \cite{BGIK-aller08,BI-rpt08} show empirically that sparse
binary measurement matrices with linear-programming (LP) reconstruction
are as good as dense random matrices. In \cite{DMM-itw101,DMM-itw102},
dense matrices with i.i.d. Gaussian random entries and an iterative
thresholding algorithm, which is a message-passing type of algorithm,
is proved to have the same sparsity-undersampling tradeoff as convex
optimization reconstruction. In \cite{LMP-aller08}, sparse measurement
matrices and message-passing decoder are used to solve a sparse signal
recovery problem in the application of per-flow data measurement on
high-speed links. %
{} All these works imply that sparse matrices with message-passing reconstruction
algorithms can be a good solution for CS systems. 

For reconstruction, the minimum number of measurements depends on
the signal model, the measurement noise, the reconstruction algorithm,
and the way reconstruction error is measured. Consider the reconstruction
of a length-$n$ signal that has $p$ non-zero (or dominant) entries.
For strictly-sparse signals, Donoho computed sparsity thresholds below
which LP reconstruction succeeds w.h.p. for high-dimensional signals
\cite{Donoho-pnas05,Donoho-dcg06}. For a compressible signal with
noisy measurements, \cite{Sarvotham-aller06} derives an information-theoretic
bound that shows $\Omega\left(p\ln(n/p)\right)$ noisy measurements
are required. In \cite{mlowerbound_crt-05}, it is shown that $O\left(p\ln(n/p)\right)$
noisy measurements are needed to reconstruct a strictly-sparse signal.
In \cite{BaIPW10}, it is shown that the lower bound $O\left(p\ln(n/p)\right)$
cannot be further improved (reduced) for a certain compressible signal
model. In this paper, we show that verification-based reconstruction
allows linear-time (in the signal dimension) reconstruction of strictly-sparse
signals with $O(p)$ measurements using real-valued measurement matrices
and noiseless measurements. At first, this seems to violate the lower
bounds on the number of measurements. However, we provide a information-theoretic
explanation that shows the $O\left(p\ln(n/p)\right)$ lower bound
does not apply to this system because the measurements are real-valued
and provide an infinite amount of information when there is no measurement
noise.

\subsection{Main Contribution}

This paper provides detailed descriptions and extensions of work reported
in two conference papers \cite{ZP-ita08,Zhang-aller08}. We believe
the main contribution of all these results are:
\begin{enumerate}
\item The observation that the Sudocodes reconstruction algorithm is an
instance of verification decoding and its decoding thresholds can
be computed precisely using numerical DE \cite{ZP-ita08}. For ensembles
with at least 3 non-zero entries in each column, this implies that
no outer code is required. For signals with $\delta n$ non-zero entries,
this reduces the lower bound on the number of noiseless measurements
required from $O(n\ln n$) to $O(n)$.
\item The introduction of the high-rate scaling analysis for iterative erasure
and verification decoding of LDPC codes \cite{ZP-ita08,Zhang-aller08}.
This technique provides closed-form upper and lower bounds on decoding
thresholds that hold uniformly as the rate approaches 1. For example,
it shows that $(3,k)$-LDPC codes achieve 81\% of capacity on the
BEC for sufficiently large $k$. This also shows that, for strictly-sparse
signals with $\delta n$ non-zero entries and noiseless measurements,
$3\delta n$ measurements are sufficient (with $(4,k)$-LDPC codes)
for verification-based reconstruction \emph{uniformly} as $\delta\rightarrow0$.
While it is known that $\delta n+1$ measurements are sufficient for
reconstruction via exhaustive search of all support sets \cite{Baron-pre05},
this shows that $O(\delta n)$ measurements also suffice for sparse
measurement matrices with low-complexity reconstruction. In constrast,
the best bounds for linear-programming reconstruction require at least
$O\left(\delta n\ln\frac{1}{\delta}\right)$ measurements.
\item The application of the high-rate scaling analysis to compute the stopping
distance of erasure and verification decoding. For example, this shows
that almost all long $(j,k)$-LDPC codes, with $j=2+\left\lceil 2\ln(k-1)\right\rceil $,
can correct all erasure patterns whose fraction of erasures is smaller
than $\frac{1}{k-1}$.
\end{enumerate}

\subsection{Structure of the Paper}

Section \ref{sec:Background} provides background information on coding
and CS. Section \ref{sec:Main-Results} summarizes the main results.
In Section \ref{sec:DE-Asymptotic-Analysis}, proofs and details are
given for the main results based on DE. While in Section \ref{sec:Stopping-Set-Asymptitic},
proofs and details are provided for the main results based on stopping-set
analysis. Section \ref{sec:sparseCSit} discusses a simple information-theoretic
bound on the number of measurements required for reconstruction. Section
\ref{sec:Simulation-Results} presents simulation results comparing
the algorithms discussed in this paper with a range of other algorithms.
Finally, some conclusions are discussed in Section \ref{sec:Conclusion}.

{[}Author's Note: The equations in this paper were originally typeset
for two-column presentation, but we have submitted it in one-column
format for easier reading. Please accept our apologies for some of
the rough looking equations.{]}

\section{\label{sec:Background} Background on Coding and CS}

\subsection{Background on LDPC Codes }

LDPC codes are linear codes introduced by Gallager in 1962 \cite{Gallager-it62}
and re-discovered by MacKay in 1995 \cite{my_ref:r2}. Binary LDPC
codes are now known to be capacity approaching on various channels
when the block length tends to infinity. They can be represented by
a Tanner graph, where the $i$-th variable node is connected to the
$j$-th check node if the entry on the $i$-th column and $j$-th
row of its parity-check matrix is non-zero. 

LDPC codes can be decoded by an iterative \emph{message-passing} (MP)
algorithm, which passes messages between the variable nodes and check
nodes iteratively. If the messages passed along the edges are probabilities,
then the algorithm is also called \emph{belief propagation} (BP) decoding.
The performance of the MP algorithm can be evaluated using density
evolution (DE) \cite{Richardson-it00} and stopping set (SS) analysis
\cite{Di-it02} \cite{orlitsky-it05}. These techniques allow one
to compute noise thresholds (below which decoding succeeds w.h.p.)
for average-case and worst-case error models, respectively.

\subsection{Encoding and Decoding}

An LDPC code is defined by its parity-check matrix $\Phi$, which
can be represented by a sparse bipartite graph. In the bipartite graph,
there are two types of nodes: variable nodes representing code symbols
and check nodes representing parity-check equations. In the standard
irregular code ensemble \cite{Luby-it01}, the connections between
variable nodes and check nodes are defined by the degree distribution
(d.d.) pairs $\lambda(x)=\sum_{i=1}^{d_{v}}\lambda_{i}x^{i-1}$ and
$\rho(x)=\sum_{i=1}^{d_{c}}\rho_{i}x^{i-1}$ where $d_{v}$ and $d_{c}$
are the maximum variable and check node degrees and $\lambda_{i}$
and $\rho_{i}$ denote the fraction of edges connected to degree-$i$
variable and check nodes, respectively. The sparse graph representation
of LDPC codes implies that the encoding and decoding algorithms can
be implemented with linear complexity in the block length\footnote{The complexity here refers to both the time and space complexity in terms of basic field operations and storage of field elements, respectively.}.
Since LDPC codes are usually defined over the finite field $GF(q)$
instead of the real numbers, we need to modify the encoding/decoding
algorithm to deal with signals over real numbers. Each entry in the
parity-check matrix is chosen either to be 0 or to be a real number
drawn from a continuous distribution. The parity-check matrix $\Phi\in\mathbb{R}^{m\times n}$
can also be used as the measurement matrix in the CS system (e.g.,
the signal vector $x\in\mathbb{R}^{n}$ is observed as $y=\Phi x$);
if there are no degree-1 nodes, then it will be full-rank with high
probability (w.h.p.). 

\begin{figure}
\centering{}\includegraphics[scale=0.5]{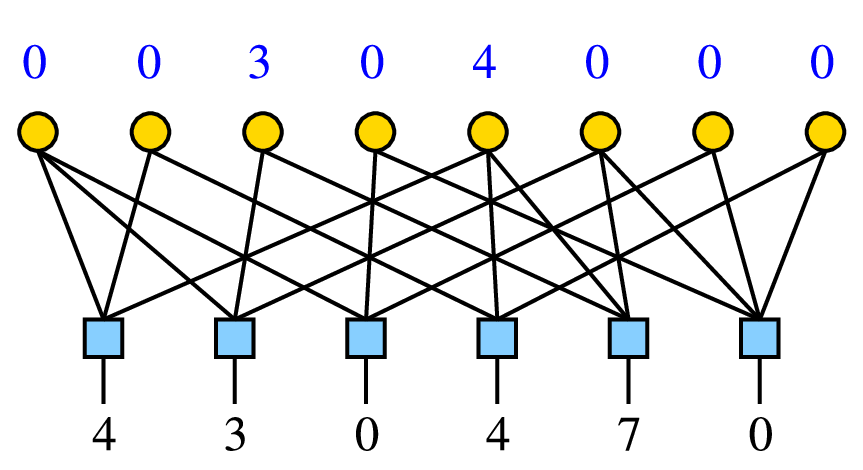}\caption{Structure of the encoder.}
\end{figure}

The process of generating the measurement variables can also be seen
from the bipartite Tanner graph representation. Figure 1 shows the
encoder structure. Each non-zero entry in $\Phi$ is the edge-weight
of its corresponding edge in this graph. Therefore, the measurement
process associated with a degree $d$ check node is as follows: 
\begin{enumerate}
\item Encoding: The measurement variable is the weighted sum (using the
edge weights) of the $d$ neighboring variable nodes given by $y_{i}=\sum_{j}\Phi_{ij}x_{i}$.
\end{enumerate}
In this work, we consider only strictly-sparse signals and we use
two decoders based on verification, which were first proposed and
analyzed in \cite{Luby-it05}. The second algorithm was also proposed
independently for CS in \cite{Sarvotham-isit06}. The decoding process
uses the following rules:
\begin{enumerate}
\item If a measurement is zero, then all neighboring variable nodes are
verified as zero. 
\item If a check node is of degree one, then verify the variable node with
the value of the measurement.
\item {[}Enhanced verification{]} If two check nodes overlap in a single
variable node and have the same measurement value, then verify that
variable node to the value of the measurement.
\item Remove all verified variable nodes and the edges attached to them
by subtracting out the verified values from the measurements.
\item Repeat steps 1-4 until decoding succeeds or makes no further progress.
\end{enumerate}
Note the first algorithm follows steps 1, 2, 4 and 5. The second algorithm
follows steps from 1 to 5. These two algorithms correspond to the
first and second algorithms in \cite{Luby-it05} and are referred
to as LM1 and node-based LM2 (LM2-NB) in this paper%
\footnote{In \cite{Luby-it05}, the second algorithm (which we refer to as LM2)
was described in a node-based (NB) fashion (as above), but analyzed
using a message-based (MB) density-evolution. There is an implicit
assumption that the two algorithms perform the same. In fact, they
perform differently and the LM2-NB algorithm is superior as observed
in \cite{Zhang-globe07}\cite{zp-it-lmp}.%
}. The Sudocodes introduced in \cite{Sarvotham-isit06} are simply
LDPC codes with a regular check d.d. and Poisson variable d.d. that
use LM2-NB reconstruction. One drawback of this choice is that the
Poisson variable d.d. with finite-mean has (w.h.p.) a linear fraction
of variables nodes that do not participate in any measurement \cite{Luby-focs02}.
For this reason, Sudocodes require a two-phase encoding that prevents
the scheme from achieving a constant oversampling rate. A detailed
discussion of the LM2-NB algorithm, which is a node-based improvement
of the message-based LM2 (LM2-MB), can be found in \cite{zp-it-lmp}.

In general, the scheme described above does not guarantee that all
verified symbols are actually correct. The event that a symbol is
verified but incorrect is called false verification (FV). In order
to guarantee there are no FVs, one can add a constraint on the signal
such that the weighted sum, of any subset of a check node's non-zero
neighbors, does not equal to zero \cite{Sarvotham-isit06,Xu-itw07}.
Another scenario where it makes sense to assume no FV is when we consider
random signals with continuous distributions so that FV occurs with
probability zero. Finally, if the measured signal is assumed to be
non-negative, then FV is impossible for the LM1 decoding algorithm. 

Verification decoding was originally introduced and analyzed for the
$q$-SC. It is based on the observation that, over large alphabets,
the probability that ``two independent random numbers are equal''
is quite small. This leads to the \emph{verification assumption} that
any two matching values (during decoding) are generated by the same
set of non-zero coefficients with high probability. The primary connection
between CS, codes over real numbers, and verification decoding lies
in the fact that:
\begin{quote}
\emph{The verification assumption applies equally well to both large
discrete alphabets and the real numbers.}
\end{quote}

\subsection{Analysis Tools}

Based on the sparse graph structure, LDPC codes can be decoded efficiently
using iterative MP algorithms. The average performance of MP decoding
algorithms can be analyzed with density evolution (DE) \cite{Richardson-it00}
or extrinsic information transfer (EXIT) charts \cite{ten-it01}.
The concentration theorem \cite{Richardson-it00} shows that random
realizations of decoding are close to the average behavior w.h.p.
for long block lengths. DE analysis provides a threshold below which
decoding (or reconstruction) succeeds w.h.p. as the block length goes
to infinity. The decoding threshold can also be improved by optimizing
the edge degree distribution (d.d.) pair $\lambda(x)$ and $\rho(x)$. 

Decoding can also be analyzed using combinatorial methods such as
stopping-set analysis \cite{Di-it02} and \cite{orlitsky-it05}. Stopping-set
analysis gives a threshold below which \emph{all} error patterns can
be recovered with certainty under the assumption of no FV. In general,
DE and stopping-set analysis lead to different thresholds. Since stopping-set
analysis implies uniform recovery of all the error patterns, instead
of just most of them, the threshold given by stopping-set analysis
is always lower than the one given by DE. For example, DE analysis
of $(3,6)$ regular codes on the BEC shows that almost all erasure
patterns of size less than $0.429$ of the block length can be corrected
w.h.p. \cite{Luby-it01}. On the other hand, stopping-set analysis
guarantees that most codes correct all erasure patterns of size less
than $0.018$ of the block length as $n\rightarrow\infty$. 

Likewise, in CS systems, there are two standard measures of reconstruction:
\emph{uniform} reconstruction and \emph{randomized} (or \emph{non-uniform})
reconstruction. A CS system achieves randomized reconstruction for
signal set (e.g., $p$-sparse signals) if \emph{most} randomly chosen
measurement matrices recover \emph{most} of the signals in the signal
set. While a CS system achieves uniform reconstruction if a measurement
matrix and the decoder recover \emph{all} the signals in the signal
set with certainty. Another criterion, which is between uniform reconstruction
and randomized reconstruction, is what we call \emph{uniform-in-probability}
reconstruction. A CS system achieves uniform-in-probability reconstruction
if, for any signal in the signal set, \emph{most} randomly chosen
measurement matrices achieve  successful decoding. 

Since DE and the concentration theorem lead to w.h.p. statements for
MP decoding over all signals and graphs, it is natural to adopt a
DE analysis to evaluate the performance of randomized reconstruction
CS systems based on LDPC codes. For uniform reconstruction, a stopping-set
analysis of the MP decoder is the natural choice. While this works
for the BEC, the possibility of FV prevents this type of strong statement
for verification decoding. If the non-zero entries of $\Phi$ are
chosen randomly from a continuous distribution, however, then the
probability of FV is zero for all signals. Therefore, one can use
stopping-set analysis to analyze MP decoding of LDPC code ensembles
and show that the LDPC codes with MP decoding achieves uniform-in-probability
reconstruction for the CS system. The reader is cautioned that these
results are somewhat brittle, however, because they rely on exact
calculation and measurement of real numbers.

While the methods discussed above can be used to numerically compute
sparsity thresholds of verification-based reconstruction for irregular
LDPC-type measurement matrices, we are particularly interested in
understanding how the number of measurements scales when the signal
is both high-dimensional and extremely sparse. To compare results,
we focus on the \emph{oversampling ratio} (i.e., the number of measurements
divided by the number of non-zero elements in the signal) required
for reconstruction. This leads us to consider the high-rate scaling
of DE and stopping-set analysis.

\subsection{Decoding Algorithms}

In CS, optimal decoding (in terms of oversampling ratio) requires
a combinatorial search that is known to be NP-Hard \cite{Candes-it05}.
Practical reconstruction algorithms tend to either be based on linear
programming (e.g., basis pursuit (BP) \cite{Chen-siamscicomp98})
or low-complexity iterative algorithms (e.g., Orthogonal Matching
Pursuit (OMP) \cite{Tropp-it07}). A wide range of algorithms allow
one to trade-off the oversampling ratio for reconstruction complexity.
In \cite{Sarvotham-isit06}, LDPC codes are used in the CS system
and the algorithm is essentially identical to the verification-based
decoding proposed in \cite{Luby-it05}. The scaling-law analysis shows
the oversampling ratio for LDPC codes based CS system can be quite
good. Encoding/decoding complexity is also a consideration. LDPC codes
have a sparse bipartite-graph representation so that encoding and
decoding is possible with complexity linear in the block length. 

There are several existing MP decoding algorithms for LDPC codes over
non-binary fields. In \cite{Luby-it01} and \cite{my_ref:shok}, an
analysis is introduced to find provably capacity-achieving codes for
erasure channels under MP decoding. Metzner presents a modified majority-logic
decoder in \cite{my_ref:Metz} that is similar to verification decoding.
Davey and MacKay develop and analyze a symbol-level MP decoder over
small finite fields \cite{my_ref:Davey}. Two verification decoding
algorithms for large discrete alphabets are proposed by Luby and Mitzenmacher
in \cite{Luby-it05} and are called LM1 and LM2 in this paper. The
list-message-passing (LMP) algorithm \cite{zp-it-lmp} provides a
smooth trade-off between the performance and complexity of the two
decoding algorithms introduced by Shokrollahi and Wang in \cite{Shokrollahi-perscomm07}.
All of these algorithms are summarized in \cite{zp-it-lmp}. 

One can get a rough idea of the performance of these algorithms by
comparing their performance for the standard $(3,6)$-regular LDPC
code. A standard performance measure is the noise threshold (or sparsity
threshold for CS) below which decoding succeeds with high probability.
The threshold of the LM1 algorithm in this case is $0.169$. This
means that a long random $(3,6)$-regular LDPC code will correct a
$q$-SC error pattern with high probability as long as the error rate
is less than $0.169$. Likewise, it means that using the same code
for LM1 reconstruction of a strictly-sparse signal will succeed w.h.p.
as long as the sparsity rate (i.e., fraction of non-zero elements)
of the signal vector is less than $0.169$. The LM2-MB algorithm improves
this threshold to $0.210$  and the LM2-NB algorithm is conjectured
to improve this threshold to $0.259$ \cite{zp-it-lmp}.

\begin{figure}
\centering{}\includegraphics[width=0.5\columnwidth]{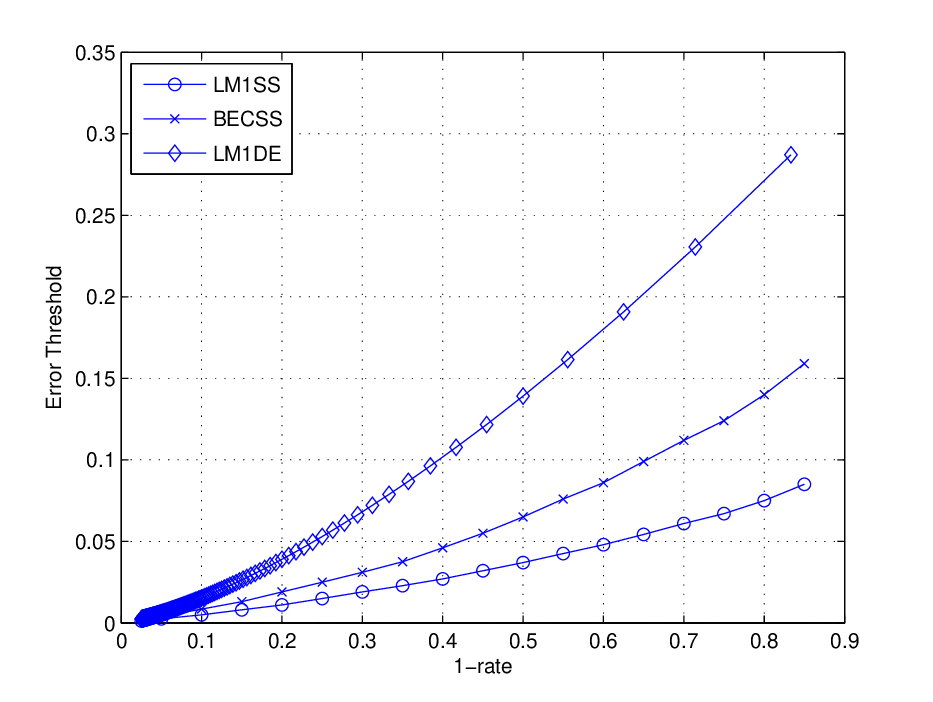}\caption{\label{fig:bounds} Thresholds vs 1-$R$, where $R$ is the code rate,
for LM1 stopping set/DE analysis and the BEC stopping-set analysis.}
\end{figure}

Likewise, the stopping-set analysis of the LM1 algorithm in Section
V shows that a $(3,6)$-regular code exists where LM1 succeeds (ignoring
FV) for all error (or sparsity) patterns whose fraction of non-zero
entries is less than $0.0055$. In comparison, the BEC stopping-set
threshold of the $(3,6)$ code is $0.018$ for erasure patterns. However,
both of these thresholds can be increased significantly (for the same
code rate) by increasing the variable node degree. In fact, the $(7,14)$-regular
LDPC code gives the best (both LM1 and BEC) stopping-set thresholds
and they are (respectively) $0.0364$ and $0.0645$. Finally, if the
signal is non-negative, then FV is not possible during LM1 decoding
and therefore $0.0364$ is a lower bound on the true LM1 rate-$\frac{1}{2}$
threshold for uniform reconstruction. Fig. \ref{fig:bounds} shows
the best decoding/recovery thresholds for regular LDPC codes with
BEC stopping-set analysis, LM1 stopping-set analysis, LM1 DE analysis
LM2-MB DE analysis and the bound by using linear programming (LP)
decoding with dense measurement matrix \cite{Donoho-pnas09}.
As we can see from Fig. \ref{fig:bounds}, LM2-MB DE gives the better
upper bound in the high-rate regime. Note that if the signal coefficients
are non-negative, the threshold of LM1 given by stopping-set analysis
is comparable to the strong bound given in \cite[Fig. 1(a)]{hassibi-it-expander},
and the threshold of LM1 given by DE analysis is comparable to the
weak bound given in \cite[Fig. 1(b)]{hassibi-it-expander}.

Since the scaling-law analysis becomes somewhat tedious when complicated
algorithms are applied, we consider only the $(j,k)$-regular code
ensemble and the relatively simple algorithms LM1 and LM2-MB. The
rather surprising result is that, even with regular codes and simple
decoding algorithms, the scaling law implies that LDPC codes with
verification decoding perform very well for noiseless CS systems with
strictly-sparse signals.

\subsection{Signal Model}

There are some significant differences between coding theory and CS.
One of them is the signal model. The first difference is that coding
theory typically uses discrete alphabets (see \cite{Wolf-com83} for
one exception to this) while CS deals with signals over the real numbers.
Fortunately, some codes designed for large discrete alphabets (e.g.,
the $q$-ary symmetric channel) can be adapted to the real numbers.
By exploring the connection and the analogy between real field and
finite field with large $q$, the CS system can be seen as an essentially
a syndrome-based source coding system \cite{ZP-ita08}. Using the
parity-check matrix of a non-binary LDPC code as the measurement matrix,
the MP decoding algorithm can be used as the reconstruction algorithm. 

The second difference in the signal model is that CS usually models
the sparse signal $x\in\mathbb{R}^{n}$ as coming from a particular
set, such as the $n$-dimensional unit $\ell_{r}$-ball. This constraint
enforces an ``approximate sparsity'' property of the signal. In
information theory and coding, the signal model is typically probabilistic.
Each component of the signal is drawn i.i.d. from a distribution,
on the real numbers, that defines the signal ensemble. A strictly-sparse
signal can be captured in this probabilistic model by choosing the
distribution to contain a Dirac delta function at zero \cite{Sarvotham-tr06,Donoho-pnas09,Baron-sp10}.

\subsection{Interesting Rate Regime}

In coding theory, the code rate depends on the application and the
interesting rate regime varies from close to zero to almost one. In
CS systems, the signal is sparse in some domain and becomes increasingly
sparse as the dimension increases. Intuitively, this means that one
can use codes with very little redundancy or very high code rate to
represent the signal. The setting where CS systems achieve the largest
gains corresponds to the high-rate regime in coding. Therefore, we
consider how the system parameters must scale as the rate goes to
one. It is important to note that the results provide bounds for a
wide range of rates, but are tight as the rate approaches one.

\section{Main Results\label{sec:Main-Results}}

The main mathematical results of this paper are now listed. Details
and proofs follow in section \ref{sec:DE-Asymptotic-Analysis} and
section \ref{sec:Stopping-Set-Asymptitic}. Note that all results
hold for asymptotically-long randomly-chosen regular LDPC codes with
variable-degree $j$ and check-degree $k$. The main idea is to fix
$j$ and observe how the decoding threshold scales when increase $k$.
This provides a scaling law for the decdoing threshold and leads to
necessary and sufficient conditions for successful reconstruction.

(i) {[}DE-BEC{]} For the BEC, there is a $K<\infty$ such that: a
check-regular LDPC code with average variable node degree $j\ge2$
and check-degree $k$ can recover a $\delta<\bar{\alpha}j/(k-1)$
fraction of erasures (w.h.p. as $n\rightarrow\infty$) when $k\ge K$.
The constant $\bar{\alpha}$ is independent of $k$ and gives the
fraction of the optimal $\delta^{*}=j/k$ threshold. Conversely, if
the erasure probability $\delta>\bar{\alpha}j/(k-1)$, then decoding
fails (w.h.p. as $n\rightarrow\infty$) for all $k$. 

(ii) {[}SS-BEC{]} For any $0\le\theta<1$, there is a $K<\infty$
such that: for all $k\geq K$, a $(j,k)$-regular LDPC code with $j\ge3$
can recover all erasures (w.h.p. as $n\rightarrow\infty$) of size
$\theta ne\left(k-1\right)^{-j/(j-2)}$. 

(iii) {[}DE-$q$-SC-LM1{]} For the $q$-SC, when one chooses a code
randomly from the $(j,k)$ regular ensemble with $j\ge2$ and uses
LM1 as decoding algorithm, then there is a $K_{1}<\infty$ such that
one can recover almost all error patterns of size $n\delta$ for $\delta<\bar{\alpha}_{j}(k-1)^{-j/(j-1)}$
(w.h.p. as $n\rightarrow\infty$) for all $k\ge K_{1}$. Conversely,
when $\delta>\bar{\alpha}_{j}(k-1)^{-j/(j-1)},$ there is a $K_{2}<\infty$
such that the decoder fails (w.h.p. as $n\rightarrow\infty$) for
all $k\ge K_{2}$. 

(iv) {[}DE-$q$-SC-LM2-MB{]} For the $q$-SC, when one chooses a code
randomly from the $(j,k)$ regular ensemble with $j\ge3$ and uses
LM2-MB as decoding algorithm, then there is a $K_{1}<\infty$ such
that one can recover almost all error patterns of size $n\delta$
for $\delta<\bar{\alpha}_{j}j/k$ (w.h.p. as $n\rightarrow\infty$).
The constant $\bar{\alpha}_{j}$ is independent of $k$ and gives
the fraction of the optimal $\delta^{*}=j/k$ threshold. Conversely,
there is a $K_{2}<\infty$ such that the decoder fails (w.h.p. as
$n\rightarrow\infty$) when $\delta>\bar{\alpha}_{j}j/k$ for all
$k\ge K_{2}$. 

(v) {[}SS-$q$-SC-LM1{]} For any $0\le\theta<1$, there is a $K<\infty$
such that: For all $k\ge K$, a $(j,k)$-regular LDPC code with $j\ge3$
using LM1 decoding can recover (w.h.p as $n\rightarrow\infty$) all
$q$-SC error patterns of size $\theta n\bar{\beta}_{j}(k-1)^{-j/(j-2)}$
if no false verifications occur.

For the sake of simplicity and uniformity, the constants $K$, $K_{1}$,
$K_{2}$ and $\bar{\alpha}_{j}$ are reused even though they may take
different values in (i), (ii), (iii), (iv) and (v).

\section{High-Rate Scaling Via Density Evolution\label{sec:DE-Asymptotic-Analysis}}

\subsection{DE Scaling-Law Analysis for the BEC}

DE analysis provides an explicit recursion, which connects the distributions
of messages passed from variable nodes to check nodes at two consecutive
iterations of MP algorithms. In the case of BEC, this DE analysis
has been derived in \cite{Luby-97} and \cite{Luby-it01}. It has
been shown that the expected fraction of erasure messages, which are
passed in the $i$-th iteration, called $x_{i}$, evolves as $x_{i}=\delta\lambda(1-\rho(1-x_{i-1}))$
where $\delta$ is the erasure probability of the channel. For general
channels, the recursion may be much more complicated because one has
to track the general distributions, which  cannot be represented by
a single parameter \cite{Richardson-it01}. 

To illustrate the scaling law, we start by analyzing the BEC case
using DE. Although this is not applicable to CS, it motivates the
scaling-law analysis for the $q$-SC, which is related to CS. 

The scaling law of LDPC codes of check-regular ensemble over the BEC
is shown by the following theorem. 
\begin{thm}
\label{thm:BEC-alpha-threshold}Consider a sequence of check-regular
LDPC codes with fixed variable degree distribution $\lambda(x)$ and
increasing check degree $k$. Let $j=1/\int_{0}^{1}\lambda(x)\mbox{d}x$
be the average variable degree and $\overline{\alpha}$, which is
called $\alpha$-threshold, be the largest $\alpha$ such that $\lambda\left(1-e^{-\alpha jx}\right)\leq x$
for $x\in[0,1]$. For the erasure probability $\delta=\alpha j/(k-1)$,
the iterative decoding of a randomly chosen length-$n$ code from
this ensemble fails (w.h.p as $n\rightarrow\infty$) for all $k$
if $\alpha>\overline{\alpha}$. Conversely, if $\alpha<\overline{\alpha}$,
then there exists a $K<\infty$ such that iterative decoding succeeds
(w.h.p as $n\rightarrow\infty$) for all $k\geq K$.
\end{thm}
Before proceeding to the proof of Theorem \ref{thm:BEC-alpha-threshold},
we introduce two lemmas that will be used throughout the paper.
\begin{lem}
\label{lem:MonotonicToExp} For all $s\geq0$, $x\ge0$ and $k^{1+s}>\left|x\right|$,
the sequence $a_{k}=\left(1-\frac{x}{k^{1+s}}\right)^{k}$ is strictly
increasing in $k$ and 
\begin{equation}
1-xk^{-s}\leq a_{k}\leq e^{-xk^{-s}}.
\end{equation}
\end{lem}
\begin{IEEEproof}
[Proof of Lemma \ref{lem:MonotonicToExp}] We restrict our attention
to $x\geq0$ because the proof is simplified in this case and the
continuation does not require $x<0$. We show that $a_{k}$ is strictly
increasing with $k$ by considering the power series expansion of
$\ln a_{k}$, which converges if $k^{1+s}>\left|x\right|$. This gives
\begin{equation}
\ln a_{k}=k\ln\left(1-\frac{x}{k^{1+s}}\right)=-xk^{-s}-\sum_{i=2}^{\infty}\frac{x^{i}}{i\, k^{(1+s)i-1}},
\end{equation}
and keeping only the first term shows that $\ln a_{k}\leq-xk^{-s}$.
Since all the terms are negative and decreasing with $k$, we see
that $a_{k}$ is strictly increasing with $k$. Since $a_{k}$ is
convex in $x$ for $k^{1+s}>\left|x\right|$, the lower bound $a_{k}\geq1-xk^{-s}$
follows from the tangent lower bound at $x=0$.
\end{IEEEproof}

\begin{lem}
\label{lem:Vitali_theorem} Let $D\subseteq\mathbb{C}$ be an open
connected set containing $[0,1]$ and $f_{k}:D\rightarrow\mathbb{C}$
be a sequence of functions that are analytic and uniformly bounded
on $D$. If $f_{k}(x)$ converges to $f_{*}(x)$ for all $x\in[0,1]$,
then $f_{k}(x)$ (all its derivatives) converge uniformly to $f_{*}(x)$
on $[0,1]$. If, in addition, $f_{k}(0)=0$ for all $k$, then $\frac{1}{x}f_{k}(x)$
also converges uniformly to $\frac{1}{x}f_{*}(x)$ on $[0,1]$. \end{lem}
\begin{IEEEproof}
[Proof of Lemma \ref{lem:Vitali_theorem}] Since the set $[0,1]\subset D$
contains an accumulation point, the first statement follows directly
from Vitali's Theorem \cite{Boas-2010}. When $f_{k}(x)=0$, it follows
from the power series about $x=0$ that the function $\frac{1}{x}f_{k}(x)$
is analytic and uniformly bounded on $D$. Therefore, Vitali's Theorem
again implies uniform convergence.
\end{IEEEproof}

\begin{IEEEproof}
[Proof of Theorem \ref{thm:BEC-alpha-threshold}] Using the substitution,
$x_{i}=\frac{\overline{\alpha}j}{k-1}y_{i}$, the DE recursion is
scaled so that 
\begin{equation}
y_{i+1}=f_{k}\left(y_{i}\right)\triangleq\frac{\alpha}{\overline{\alpha}}\lambda\left(1-\left(1-\frac{\overline{\alpha}jy_{i}}{k-1}\right)^{k-1}\right).\label{eq:ScaledBECRecursion}
\end{equation}
By Lemma \ref{lem:MonotonicToExp}, $(1-\frac{x}{k-1})^{k-1}$ increases
monotonically (for $x\leq k-1$) to $e^{-x}$, and therefore $f_{k}(y)$
decreases monotonically to $f_{*}(y)=\frac{\alpha}{\overline{\alpha}}\lambda\left(1-e^{-\overline{\alpha}jy}\right)$.
If $\alpha>\overline{\alpha}$, then the definition of $\overline{\alpha}$
implies that $f_{*}(y')>y'$ for some $y'\in[0,1]$. Since $f_{k}(y')\geq f_{*}(y')>y'$
and each $f_{k}(y)$ is continuous, it follows that the recursion
$y_{i+1}=f_{k}\left(y_{i}\right)$ will not converge to zero (from
$y_{0}=1$) for all $k\geq2$. Therefore, iterative decoding will
also fail w.h.p. as $n\rightarrow\infty$. 

For the next part, we notice that each $f_{k}(y)$ is an entire function
satisfying $f_{k}(0)=0$ and $\left|f_{k}(y)\right|\leq\frac{\alpha}{\overline{\alpha}}\lambda\left(1+e^{\overline{\alpha}j|y|}\right)$.
Therefore, we can apply Lemma \ref{lem:Vitali_theorem} (with $D=\left\{ y\in\mathbb{C}\:|\:\left|y\right|\leq2\right\} $)
to see that $\frac{1}{y}f_{k}(y)$ is a sequence of continuous functions
that converges uniformly to $\frac{1}{y}f_{*}(y)$ on $[0,1]$. If
$\alpha<\overline{\alpha}$, then the definition of $\overline{\alpha}$
implies that $\frac{1}{y}f_{*}(y)\leq\frac{\alpha}{\overline{\alpha}}$
for $y\in[0,1]$. Since $\frac{1}{y}f_{k}(y)\searrow\frac{1}{y}f_{*}(y)$
uniformly on $[0,1]$, there must exist a $K<\infty$ such that $\frac{1}{y}f_{k}(y)\leq\frac{\alpha+\overline{\alpha}}{2\overline{\alpha}}$
for $y\in[0,1]$ and $k\geq K$. Therefore, the recursion $y_{i+1}=f_{k}\left(y_{i}\right)$
will converge to zero (from $y_{0}=1$) for all $k\geq K$ and iterative
decoding will succeed w.h.p. as $n\rightarrow\infty$. In practice,
one can choose $K$ to be the smallest $k$ such that $f_{k}(y)<y$
for $y\in(0,1]$. 
\end{IEEEproof}
The following corollary determines a few $\alpha$-thresholds explicitly.
\begin{cor}
\label{cor:BEC-alpha-thr}For $(j,k)$ regular LDPC codes, the $\alpha$-threshold
of BEC decoding is given by $\overline{\alpha}_{j}$ with $\overline{\alpha}_{2}=0.5$,
$0.8184<\overline{\alpha}_{3}<0.8185$, and $0.7722<\overline{\alpha}_{4}<0.7723$.\end{cor}
\begin{IEEEproof}
See Appendix \ref{sec:Appendix1}.\end{IEEEproof}
\begin{rem}
For example, if $j=3$ and $\alpha=0.75<\overline{\alpha}_{3}$, then
numerical results show that $K=9$ suffices so that DE converges for
all $k\geq9$ when $\delta<3(0.75)/(k-1)$. Therefore, this approach
provides a lower bound on the threshold for all $k\geq9$ that is
tight as $k\rightarrow\infty$.
\end{rem}

\subsection{DE Scaling-Law Analysis for the $q$-SC}

\subsubsection{DE Scaling-Law Analysis for LM1}

For the simplicity of our analysis, we only consider $(j,k)$-regular
code ensemble and the LM1 decoding algorithm \cite{Luby-it05} for
the $q$-SC with error probability $\delta$. The DE recursion for
LM1 is (from \cite{Luby-it05})
\begin{equation}
\!\; x_{i+1}\!=\!\delta\left(\!1\!-\!\left[1\!-\!\left(1\!-\!\delta\right)\left(1\!-\!(1\!-\! x_{i})^{k-1}\right)^{j-1}\!\!\!\!\!\!-x_{i}\right]^{k-1}\right)^{j-1}\!\!\!\!\!\!\!\!\!,\label{eq:de_lm1}
\end{equation}
where $x_{i}$ is the fraction of unverified messages in the $i$-th
iteration. Our analysis of the scaling law relies on the following
lemma. 
\begin{lem}
\label{lem:Let-the-functions}Let the functions $g_{k+1}(x)$ and
$\overline{g}_{k+1}(x)$ be defined by \vspace{-7mm} 
\begin{multline*}
\\
g_{k+1}(x)\triangleq\frac{\alpha}{\overline{\alpha}_{j}}\Biggl(1-\Biggl[1-\left(1-\frac{\alpha}{k^{j/(j-1)}}\right)\\
\biggl(1-\Bigl(1-\frac{\overline{\alpha}_{j}x}{k^{j/(j-1)}}\Bigr)^{k}\biggr)^{j-1}\!\!\!\!\!-\frac{\overline{\alpha}_{j}x}{k^{j/(j-1)}}\Biggr]^{k}\Biggr)^{j-1}\\
\vspace{-10mm}
\end{multline*}
and

\[
\overline{g}_{k+1}(x)\triangleq\frac{\alpha}{\overline{\alpha}_{j}}\left(1-\left[1-\frac{\overline{\alpha}_{j}^{j-1}x^{j-1}}{k}-\frac{\overline{\alpha}_{j}x}{k^{j/(j-1)}}\right]^{k}\right)^{j-1}\!\!\!\!\!\!,
\]
where $\overline{\alpha}_{j}\geq1$, $\alpha\in(0,\overline{\alpha}_{j}]$,
and $j\geq2$. For $x\in(0,1]$ and $k>\overline{\alpha}_{j}^{j-1}$,
these functions satisfy (i) $g_{k}(x)\leq\overline{g}_{k}(x)$, (ii)
$\overline{g}_{k}(x)$ is monotonically decreasing with $k$ for $k>\overline{\alpha}_{j}^{j-1}$,
and (iii) $g_{*}(x)\triangleq\lim_{k\rightarrow\infty}g_{k}(x)=\lim_{k\rightarrow\infty}\overline{g}_{k}(x)=\frac{\alpha}{\overline{\alpha}_{j}}\left(1-e^{-\overline{\alpha}_{j}^{j-1}x^{j-1}}\right)^{j-1}$.\end{lem}
\begin{IEEEproof}
See the Appendix \ref{sec:Appendix2}.\end{IEEEproof}
\begin{thm}
\label{thm:thm2} Consider a sequence of $(j,k)$-regular LDPC codes
with fixed variable degree $j\ge2$ and increasing check degree $k$.
Let $\overline{\alpha}_{j}$ be the largest $\alpha$ such that $(1-e^{-\alpha^{j-1}x^{j-1}})^{j-1}\leq x$
for $x\in[0,1]$. If the sparsity of the signal is $n\delta$ for
$\delta=\alpha(k-1)^{-j/(j-1)}$ and $\alpha<\overline{\alpha}_{j}$,
then there exists a $K_{1}$ such that by randomly choosing a length-$n$
code from the $(j,k)$ regular LDPC code ensemble, LM1 reconstruction
succeeds (w.h.p as $n\rightarrow\infty$) for all $k\geq K_{1}$.
Conversely, if $\alpha>\overline{\alpha}_{j}$ then there exists a
$K_{2}$ such that LM1 reconstruction fails (w.h.p as $n\rightarrow\infty$)
for all $k\ge K_{2}$.\end{thm}
\begin{IEEEproof}
Scaling  (\ref{eq:de_lm1}) using the change of variables $\delta=\alpha(k-1)^{-j/(j-1)}$
and $x_{i}=\overline{\alpha}_{j}y_{i}(k-1)^{-j/(j-1)}$ gives $y_{i+1}=g_{k}\left(y_{i}\right)$.
Lemma \ref{lem:Let-the-functions} defines the sequence $\overline{g}_{k}(y)$
and shows that $g_{k}(y)\leq\overline{g}_{k+1}(y)\leq\overline{g}_{k}(y)$
for $k>\overline{\alpha}_{j}^{j-1}$. It will also be useful to observe
that $\frac{1}{y}g_{k}(y)$ and $\frac{1}{y}\overline{g}_{k}(y)$
are both sequences of continuous functions that converge uniformly
to $\frac{1}{y}g_{*}(y)$ on $[0,1]$. To see this, we can apply Lemma
\ref{lem:Vitali_theorem} with $D=\left\{ y\in\mathbb{C}\:|\:\left|y\right|\leq2\right\} $
because $g_{k}(y)$ and $\overline{g}_{k}(y)$ are sequences of entire
functions that can be uniformly bounded on $D$.

If $\alpha<\overline{\alpha}_{j}$, then the definition of $\overline{\alpha}_{j}$
implies that $\frac{1}{y}g_{*}(y)\leq\frac{\alpha}{\overline{\alpha}_{j}}$
for all $y\in[0,1]$. Since $\frac{1}{y}\overline{g}_{k}(y)\searrow\frac{1}{y}g_{*}(y)$
uniformly on $[0,1]$, there must exist a $K_{1}<\infty$ such that
$\frac{1}{y}\overline{g}_{k}(y)\leq\frac{\alpha+\overline{\alpha}}{2\overline{\alpha}}$
for $y\in[0,1]$ and $k\geq K_{1}$. Since $g_{k}(y)\leq\overline{g}_{k}(y)$,
the recursion $y_{i+1}=g_{k}\left(y_{i}\right)$ will converge to
zero (from $y_{0}=1$) for all $k\geq K_{1}$ and iterative decoding
will succeed w.h.p. as $n\rightarrow\infty$. In practice, one can
choose $K_{1}<\infty$ to be the smallest $k$ such that $\overline{g}_{k}(y)<y$
for all $y\in(0,1]$.

If $\alpha>\overline{\alpha}_{j}$, then (by the definition of $\overline{\alpha}_{j}$)
$g_{*}(y')>y'$ for some $y'\in[0,1]$. Since $\lim_{k\rightarrow\infty}g_{k}(y)=g_{*}(y)$
, there must exist a $K_{2}$ such that $g_{k}(y')>y'$ for all $k\geq K_{2}$.
Since each $g_{k}(y)$ is continuous, the recursion $y_{i+1}=g_{k}\left(y_{i}\right)$
will not converge to zero (from $y_{0}=1$) and iterative decoding
will fail w.h.p. as $n\rightarrow\infty$ for all $k\geq K_{2}$. \end{IEEEproof}
\begin{rem}
Consider a randomly chosen code from the $(j,k)$ regular ensemble
is applied to a CS system with LM1 reconstruction. For sufficiently
large $k$, randomized reconstruction succeeds (w.h.p as $n\rightarrow\infty$)
when the sparsity is $n\delta$ with $\delta<\delta_{0}\triangleq\bar{\alpha}_{j}(k-1)^{-j/(j-1)}$.
Let $\gamma_{0}\triangleq\frac{j}{\delta_{0}(k-1)}=\bar{\alpha}_{j}^{-(j-1)/j}\delta_{0}^{-1/j}j$
and observe that an oversampling ratio $\gamma=\frac{j}{\delta k}$
larger than $\gamma_{0}$ implies $\delta<\frac{k-1}{k}\delta_{0}$.
This implies that $m=\gamma n\delta$ measurements suffice (w.h.p
as $n\rightarrow\infty$) for $\gamma>\bar{\alpha}_{j}^{-(j-1)/j}\delta_{0}^{-1/j}j$
and sufficiently small $\delta_{0}$.

The following lemma shows how to calculate the scaled threshold $\bar{\alpha}_{j}$.\end{rem}
\begin{cor}
\label{lem:lm1_alpha} For $\left(j,k\right)$ regular LDPC codes
with $j\geq2$, the $\alpha$-threshold of LM1-MB is given by $\bar{\alpha}_{j}\geq1$
and numerical calculations show $\bar{\alpha}_{2}=1$, $1.8732<\bar{\alpha}_{3}<1.8733,$
$1.6645<\bar{\alpha}_{4}<1.6646$ and $1.5207<\bar{\alpha}_{5}<1.5208$.\end{cor}
\begin{IEEEproof}
See Appendix \ref{sec:Appendix3}.\end{IEEEproof}
\begin{cor}
For regular LDPC codes and LM1 reconstruction, choosing $j=\left\lceil \ln\frac{1}{\delta}\right\rceil $
allows one to upper bound the oversampling ratio by $\left\lceil \ln\frac{1}{\delta}\right\rceil e$
for sufficiently small $\delta$.\end{cor}
\begin{IEEEproof}
For sufficiently small $\delta$, a sufficient oversampling ratio
is $\gamma_{0}=\bar{\alpha}_{j}^{-(j-1)/j}j\delta^{-1/j}\le j\delta^{-1/j}$
because $\bar{\alpha}_{j}\geq1$. Choosing $j=\left\lceil \ln\frac{1}{\delta}\right\rceil $
and taking the logarithm of both sides shows that 
\begin{equation}
\ln\gamma_{0}\le\ln\left\lceil \ln\frac{1}{\delta}\right\rceil +\frac{1}{\left\lceil \ln\frac{1}{\delta}\right\rceil }\ln\frac{1}{\delta}\le\ln\left\lceil \ln\frac{1}{\delta}\right\rceil +1.
\end{equation}

\end{IEEEproof}

\subsubsection{Scaling-Law Analysis Based on DE for LM2-MB}

For the second algorithm in \cite{Luby-it05}, the DE recursion for
the fraction $x_{i}$ of unverified messages in the $i$-th iteration
is \vspace{-7mm} 
\begin{multline}
\\
x_{i+1}=\delta\biggl(\lambda\left(1-\rho\left(1-x_{i}\right)\right)+\lambda^{'}\left(1-\rho\left(1-x_{i}\right)\right)\\
\biggl(\rho\left(1-x_{i}\right)-\rho\Bigl(1-\left(1-\delta\right)\lambda\left(1-\rho\left(1-x_{i}\right)\right)-x_{i}\Bigr)\biggr)\biggr).\\
\vspace{-10mm}\label{eq:LM2DE}
\end{multline}
Like the analysis of LM1, we first introduce a lemma to bound the
scaled DE equation.
\begin{lem}
\label{lem:LM2_bound}The functions $g_{k}(x)$ and $\bar{g}_{k}(x)$
are defined as\vspace{-7mm}

{\small 
\begin{multline*}
\\
\!\!\!\! g_{k}(x)\triangleq\frac{\alpha}{\bar{\alpha}_{j}}\Biggl(\Bigl(s(x)\Bigr)^{j-1}+\left(j-1\right)\Bigl(s(x)\Bigr)^{j-2}\left(1-\frac{\alpha jx}{k}\right)^{k-1}\\
\biggl(1-\biggl(1-\frac{1-\frac{\alpha j}{k}}{1-\frac{\alpha jx}{k}}\Bigl(s(x)\Bigr)^{j-1}\biggr)^{k-1}\biggr)\Biggr),\\
\vspace{-10mm}
\end{multline*}
}where $s(x)=1-\left(1-\frac{\alpha jx}{k}\right)^{k-1}$, i.e., $1-\rho(1-y)$,
and\vspace{-7mm}

\begin{multline*}
\\
\bar{g}_{k}(x)\triangleq\frac{\alpha}{\bar{\alpha}_{j}}\Biggl(\left(1-\left(1-\frac{\alpha jx}{k}\right)^{k}\right)^{j-1}\\
+\left(j-1\right)\left(1-\left(1-\frac{\alpha jx}{k}\right)^{k}\right)^{j-2}\left(1-\frac{\alpha jx}{k}\right)^{k}\Biggr).\\
\vspace{-10mm}
\end{multline*}
For $x\in(0,1]$ and $k>\alpha$, these functions satisfy (i) $g_{k}(x)\leq\bar{g}_{k}(x)$,
(ii) $\lim_{k\rightarrow\infty}g_{k}(x)=\lim_{k\rightarrow\infty}\overline{g}_{k}(x)=g_{*}(x)$
where 
\begin{equation}
g_{*}(x)\triangleq\frac{\alpha}{\bar{\alpha}_{j}}\left(1-e^{-\alpha jx}\right)^{j-2}\left(1+(j-2)e^{-\alpha jx}\right),\label{eq:limit_g_lm2}
\end{equation}
and (iii) $\bar{g}_{k}(x)$ is a monotonically decreasing function
of $k$. \end{lem}
\begin{IEEEproof}
See the Appendix \ref{sec:Appendix4}.\end{IEEEproof}
\begin{thm}
\label{thm:Consider-a-sequence}Consider a sequence of $(j,k)$-regular
LDPC codes with variable node degree $j\ge3$. Let $\overline{\alpha}_{j}$
be the largest $\alpha$ such that $\left(1-e^{-\alpha jx}\right)^{j-2}\left(1+(j-2)e^{-\alpha jx}\right)\leq x$
for $x\in[0,1]$. If the sparsity of the signal is $n\delta$ with
$\delta=\alpha j/k$ and $\alpha<\overline{\alpha}_{j}$, then there
exists a $K_{1}$ such that LM2-MB reconstruction succeeds (w.h.p
as $n\rightarrow\infty$) for all $k\geq K_{1}$. Conversely, if $\alpha>\overline{\alpha}_{j}$
then there exists a $K_{2}$ such that LM2-MB decoding fails (w.h.p
as $n\rightarrow\infty$) for all $k\ge K_{2}$ .\end{thm}
\begin{IEEEproof}
The LM2-MB DE recursion is given by (\ref{eq:LM2DE}). Using the change
of variables $x_{i}=\frac{\overline{\alpha}_{j}j}{k}y_{i}$ and $\delta=\frac{\alpha j}{k}$,
the scaled DE equation can be written as $y_{i+1}=g_{k}(y_{i}).$
Lemma \ref{lem:LM2_bound} defines the sequence $\overline{g}_{k}(y)$
and shows that $g_{k}(y)\leq\overline{g}_{k+1}(y)\leq\overline{g}_{k}(y)$.
It will also be useful to observe that $\frac{1}{y}g_{k}(y)$ and
$\frac{1}{y}\overline{g}_{k}(y)$ are both sequences of continuous
functions that converge uniformly to $\frac{1}{y}g_{*}(y)$ on $[0,1]$.
To see this, we can apply Lemma \ref{lem:Vitali_theorem} with $D=\left\{ y\in\mathbb{C}\:|\:\left|y\right|\leq2\right\} $
because $g_{k}(y)$ and $\overline{g}_{k}(x)$ are sequences of entire
functions that are uniformly bounded on $D$.

If $\alpha<\overline{\alpha}_{j}$, then the definition of $\bar{\alpha}_{j}$
implies that $\frac{1}{y}g_{*}(y)<\frac{\alpha}{\overline{\alpha}_{j}}$
for $y\in[0,1]$. Since $\overline{g}_{k}(y)\searrow g_{*}(y)$ uniformly
on $[0,1]$, there must exist a $K_{1}<\infty$ such that $\frac{1}{y}\overline{g}_{k}(y)\leq\frac{\alpha+\overline{\alpha}}{2\overline{\alpha}}$
for $y\in[0,1]$ and $k\geq K_{1}$. Since $g_{k}(y)\leq\overline{g}_{k}(y)$,
the recursion $y_{i+1}=g_{k}\left(y_{i}\right)$ will converge to
zero (from $y_{0}=1$) for all $k\geq K_{1}$ and iterative decoding
will succeed w.h.p. as $n\rightarrow\infty$. In practice, one can
choose $K_{1}<\infty$ to be the smallest $k$ such that $\overline{g}_{k}(y)<y$
for all $y\in(0,1]$.

If $\alpha>\overline{\alpha}_{j}$, then (by the definition of $\overline{\alpha}_{j}$)
$g_{*}(y')>y'$ for some $y'\in[0,1]$. Therefore, there exists a
$K_{2}<\infty$ such that $g_{k}(y')>y'$ for all $k\ge K_{2}$. Since
each $g_{k}(y)$ is continuous, the recursion $y_{i+1}=g_{k}\left(y_{i}\right)$
does not converge to zero (from $y_{0}=1$) and iterative decoding
will fail w.h.p. as $n\rightarrow\infty$ for all $k\geq K_{2}$.

For $j=2$, the quantity $\overline{\alpha}_{2}$ is undefined because
$\left(1-e^{-\alpha jx}\right)^{j-2}\left(1+(j-2)e^{-\alpha jx}\right)=1$.
This implies that $(2,\, k)$ regular LDPC codes do not obey this
scaling law for LM2-MB decoding.\end{IEEEproof}
\begin{rem}
If a randomly chosen code from the $(j,k)$ regular ensemble is applied
to a CS system with LM2-MB reconstruction, then randomized reconstruction
succeeds (w.h.p as $n\rightarrow\infty$) when the sparsity is $n\delta$
with $\delta<\bar{\alpha}_{j}j/k$. This requires $m\ge\gamma n\delta$
measurements and an oversampling ratio of $\gamma>\gamma_{0}=1/\bar{\alpha}_{j}$.
\end{rem}
\begin{rem}
For $\left(j,k\right)$ regular LDPC codes, the $\alpha$-threshold
of LM2-MB is given by $\overline{\alpha}_{j}$ and can be calculated
numerically to get $\overline{\alpha}_{3}=\frac{1}{6},$ $0.3416<\overline{\alpha}_{4}<0.3417$
and $0.3723<\overline{\alpha}_{5}<0.3724$.
\end{rem}
The interesting part of this result is that the number of measurements
needed for randomized reconstruction with LM2-MB (as $n\to\infty$)
is upper bounded by $\gamma\delta n$ \emph{uniformly} as $\delta\rightarrow0$.
All other reconstruction methods with moderate complexity require
$O\left(\delta n\ln\frac{1}{\delta}\right)$ measurements as $\delta\to0$.

\section{Scaling Laws Based on Stopping-Set Analysis\label{sec:Stopping-Set-Asymptitic}}

DE analysis provides the threshold below which the \emph{randomized}
(or \emph{non-uniform}) recovery is guaranteed, in the following sense:
the signal and the measurement matrix are both chosen randomly, and
w.h.p. the reconstruction algorithm gives the correct answer. If the
reconstruction algorithm is guaranteed to succeed for all signals
of sufficient sparsity, this is called \emph{uniform} recovery. On
the other hand, if reconstruction algorithm is uniform over all support
sets of sufficient sparsity, but succeeds w.h.p. over the amplitudes
of the non-zero elements (i.e., has a small but non-zero failure probability
based on amplitudes), then the reconstruction is called \emph{uniform-in-probability}
recovery. 

According to the analysis in section \ref{sec:DE-Asymptotic-Analysis},
we know that the number of measurements needed for randomized recovery
by using LM2-MB is $O(p)$ for a $p$-sparse signal. Still, the reconstruction
algorithm may fail due to the support set (e.g., it reaches a stopping
set) or due to the non-zero amplitudes of the signal  (e.g., a false
verification occurs).

In this section, we will analyze the performance of MP decoding algorithms
with uniform-in-probability recovery in the high-rate regime. This
follows from a stopping-set analysis of the decoding algorithms. A
stopping set is defined as an erasure pattern (or internal decoder
state) from which the decoding algorithm makes no further progress.
Following the definition in \cite{Di-it02}, we let $G=(V\cup C,E)$
be the Tanner graph of a code where $V$ is the set of variable nodes,
$C$ is the set of check nodes and $E$ is the set of edges between
$V$ and $C.$ A subset $U\subseteq V$ is a BEC stopping set if no
check node is connected to $U$ via a single edge. The scaling law
below uses the average stopping-set enumerator for LDPC codes as a
starting point.

\subsection{Scaling-Law Analysis for Stopping Sets on the BEC}

The \emph{average stopping set distribution} $E_{n,j,k}(s)$ is defined
as the average (over the ensemble) number of stopping sets with size
$s$ in a randomly chosen $(j,k)$ regular code with $n$ variable
nodes. The \emph{normalized stopping set distribution} $\gamma_{j,k}(\alpha)$
is defined as $\gamma_{j,k}(\alpha)\triangleq\lim_{n\rightarrow\infty}\frac{1}{n}\ln E_{n,j,k}(n\alpha).$
The \emph{critical stopping ratio} $\alpha_{j,k}^{*}$ is defined
as $\alpha_{j,k}^{*}\triangleq\inf\{\alpha>0:\gamma_{j,k}(\alpha)\ge0\}.$
Intuitively, if the normalized size of a stopping set is greater than
or equal to $\alpha_{j,k}^{*},$ then the average number of stopping
sets grows exponentially with $n.$ If the normalized size is less
than $\alpha_{j,k}^{*},$ then the average number of stopping sets
decays exponentially with $n$. In fact, there exist codes with no
stopping sets of normalized size less than $\alpha_{j,k}^{*}$. Therefore,
the quantity $\alpha_{j,k}^{*}$ can also be thought of as a \emph{deterministic
decoding threshold.} 

The normalized average stopping set distribution $\gamma_{j,k}(\alpha)$
for $(j,k)$ regular ensembles on the BEC is given by \cite{orlitsky-it05}
\[
\gamma_{j,k}(\alpha)\!\le\!\gamma_{j,k}(\alpha;x)\!\triangleq\!\frac{j}{k}\ln\left(\frac{\left(1\!+\! x\right)^{k}\!-\! kx}{x^{k\alpha}}\right)\!-\!(j\!-\!1)h(\alpha),
\]
where $h(\cdot)$ is the entropy of a binary distribution and the
bound holds for any $0\leq x\leq1$. The optimal value $x_{0}$ is
the unique positive solution of 
\begin{equation}
\frac{x((1+x)^{k-1}-1)}{(1+x)^{k}-kx}=\alpha.\label{eq:BEC_ss_x0}
\end{equation}
This gives the following theorem.
\begin{thm}
For any $0\le\theta<1$, there is a $K<\infty$ such that, for all
$k\ge K$, a randomly chosen $(j,k)$ regular LDPC code ($j\geq3$)
will (w.h.p. as $n\rightarrow\infty$) correct all erasure patterns
of size less than $\theta ne(k-1)^{-j/(j-2)}$. \end{thm}
\begin{IEEEproof}
[Sketch of Proof] Here, we provide a sketch of proof for the interest
of brevity. Since there is no explicit solution for $x_{0}$, we use
a 2nd order expansion of the LHS of  (\ref{eq:BEC_ss_x0}) around
$x=0$ and solve for $x$. This gives $x_{0}=\sqrt{\frac{\alpha}{k-1}}+o(\alpha)$.
Since $\gamma_{j,k}(\alpha)\le\gamma_{j,k}(\alpha,x)$ holds for all
$x\ge0$, we have
\begin{equation}
{\textstyle \!\gamma_{j,k}(\alpha)\!\le\!\frac{j}{k}\ln\left(\frac{\left(1+\sqrt{\frac{\alpha}{k-1}}\right)^{k}-k\sqrt{\frac{\alpha}{k-1}}}{\left(\frac{\alpha}{k-1}\right)^{\frac{k\alpha}{2}}}\right)\!-\!(j-1)h(\alpha).}\label{eq:scaling_bec_gamma}
\end{equation}
Next we expand the RHS of (\ref{eq:scaling_bec_gamma}) around $\alpha=0$
and neglect the high order terms; solving for $\alpha$ gives an upper
bound on the critical stopping ratio
\[
\alpha_{j,k}^{*}\ge e(k-1)^{-j/(j-2)}.
\]
It can be shown that this bound on $\alpha_{j,k}^{*}$ is tight as
$k\rightarrow\infty$. This means that, for any $0\le\theta<1$, there
is a $K$ such that $\theta e(k-1)^{-j/(j-2)}\le\alpha_{j,k}^{*}\le e(k-1)^{-j/(j-2)}$
for all $k>K$. Therefore, the critical stopping ratio $\alpha_{j,k}^{*}$
scales like $e(k-1)^{-j/(j-2)}$ as $k\rightarrow\infty$.\end{IEEEproof}
\begin{rem}
Although the threshold is strictly increasing with $j$, this ignores
the fact that the code rate is decreasing with $j$. However, if one
optimizes the oversampling ratio instead, then the choice of $j^{*}=2+\left\lceil 2\ln(k-1)\right\rceil $
is nearly optimal. Moreover, it leads to the simple result $\alpha_{j^{*},k}^{*}\geq\frac{1}{k-1}$
which implies an oversampling ratio that grows logarithmically in
$k$. In fact, this oversampling ratio is only a factor of 2 larger
than the optimal result implied by the binary entropy function.
\end{rem}

\subsection{Stopping-Set Analysis for the $q$-SC with LM1-NB}

A stopping set for LM1-NB is defined by considering a decoder where
$S,T,U$ are disjoint subsets of $V$ corresponding to verified, correct,
and incorrect variable nodes. Decoding progresses if and only if (i)
a check node has all but one edge attached to $S$ or (ii) a check
node has all edges attached to $S\cup T$. Otherwise, the pattern
is a stopping set. In the stopping-set analysis for $q$-SC, we can
define $E_{n,j,k}(\alpha,\beta)$ as the \emph{average number of stopping
sets} with $\left|T\right|=n\alpha$ correctly received variable nodes
and $\left|U\right|=n\beta$ incorrectly received variable nodes where
$n$ is the code length. 

The average number of stopping sets $E_{n,j,k}(\alpha,\beta)$ can
be computed by counting the number of ways, $S_{n,j,k}(a,b)$, that
$a$ correct variable nodes, $b$ incorrect variables nodes, and $n-a-b$
verified variable nodes can be connected to $\frac{nj}{k}$ check
nodes to form a stopping set. The number $S_{n,j,k}(a,b)$ can be
computed using the generating function for one check, 
\[
g_{k}(x,y)\triangleq(1+x+y)^{k}-ky-\left((1+x)^{k}-1\right),
\]
which enumerates the number of edge connection patterns (``1'' counts
verified edges, ``$x$'' counts correct edges, and ``$y$'' counts
incorrect edges) that prevent decoder progress. Generalizing the approach
of \cite{orlitsky-it05} gives 
\begin{equation}
E_{n,j,k}(\alpha,\beta)\!=\!\frac{{\scriptstyle \left(\begin{array}{c}
n\\
n\alpha,n\beta,n(1-\alpha-\beta)
\end{array}\right)}S_{n,j,k}(\alpha n,\beta n)}{{\scriptstyle \left(\begin{array}{c}
nj\\
nj\alpha,nj\beta,nj(1-\alpha-\beta)
\end{array}\right)}}\label{eq:average_number_ss}
\end{equation}
where 
\[
S_{n,j,k}(a,b)\triangleq\mbox{coeff}\left(g_{k}(x,y)^{nj/k},x^{ja}y^{jb}\right).
\]

For this work, we are mainly interested in largest $\beta$ for which
$E_{n,j,k}(\alpha,\beta)$ goes to zero as $n\rightarrow\infty$.
Since the growth (or decay) rate of $E_{n,j,k}(\alpha,\beta)$ is
exponential in $n$, this leads us to consider the \emph{normalized
average stopping set distribution} $\gamma_{j,k}(\alpha,\beta)$,
which is defined as 
\begin{equation}
\gamma_{j,k}(\alpha,\beta)=\lim_{n\to\infty}\frac{1}{n}\ln E_{n,j,k}(\alpha,\beta).
\end{equation}
Likewise, the \emph{critical stopping ratio} $\beta_{j,k}^{*}$ is
defined as 
\begin{equation}
\beta_{j,k}^{*}=\inf\{\beta\in[0,1]:w_{j,k}(\beta)>0\}
\end{equation}
where 
\[
w_{j,k}(\beta)\triangleq\sup_{\alpha\in[0,1-\beta]}\gamma_{j,k}(\alpha,\beta).
\]
Note that $w_{j,k}(\beta)$ describes the asymptotic growth rate of
the average number of stopping sets with number of incorrectly received
nodes $n\beta.$ The average number of stopping sets with size less
than $n\beta_{j,k}^{*}$ decays exponentially with $n$ and the ones
with size larger than $n\beta_{j,k}^{*}$ grows exponentially with
$n.$
\begin{thm}
The normalized average stopping set distribution $\gamma_{j,k}(\alpha,\beta)$
for LM1 can be bounded by

\vspace{-12mm}
\begin{multline}
\\
\gamma_{j,k}(\alpha,\beta)\le\gamma_{j,k}(\alpha,\beta;x,y)\triangleq\\
\frac{j}{k}\ln\frac{\left(1+(1+x+y)^{k}-ky-(1+x)^{k}\right)}{x^{k\alpha}y^{k\beta}}\\
+(1-j)h(\alpha,\beta,1-\alpha-\beta)\\
\vspace{-10mm}\label{eq:gamma}
\end{multline}
where the tightest bound is given by choosing $(x,y)$ to be the unique
positive solution of 
\begin{equation}
\frac{x\left(\left(1+x+y\right)^{k-1}-(1+x)^{k-1}\right)}{1+\left(1+x+y\right)^{k}-ky-(1+x)^{k}}=\alpha\label{eq:x0}
\end{equation}
 and
\begin{equation}
\frac{y\left(\left(1+x+y\right)^{k-1}-1\right)}{1+\left(1+x+y\right)^{k}-ky-(1+x)^{k}}=\beta.\label{eq:y0}
\end{equation}
\end{thm}
\begin{IEEEproof}
Starting from  (\ref{eq:average_number_ss}) and using Stirling's
formula, it can be verified easily that 
\[
\lim_{n\rightarrow\infty}\!\frac{1}{n}\!\ln\!\frac{{\textstyle \binom{n}{n\alpha,n\beta,n(1-\alpha-\beta}}}{{\textstyle \binom{nj}{nj\alpha,nj\beta,nj(1-\alpha-\beta}}}\!=\!(1-j)h(\alpha,\beta,1-\alpha-\beta),
\]
where $h(\cdot)$ is the entropy of a ternary distribution. Using
a Chernoff-type bound for $S_{n,j,k}(a,b)$ (i.e., $\mbox{coeff}\left(f(x,y),x^{i}y^{j}\right)\le\frac{f(x,y)}{x^{i}y^{j}}$
for all $x,y>0$), we define 
\[
\psi_{j,k}(\alpha,\beta;x,y)\!\triangleq\!\frac{nj}{k}\ln\frac{\left(1\!+\!(1\!+\! x\!+\! y)^{k}\!-\! ky\!-\!(1\!+\! x)^{k}\right)}{x^{k\alpha}y^{k\beta}}.
\]
Minimizing the bound over $x,y$ gives 
\begin{align*}
\gamma_{j,k}(\alpha,\beta) & \le\gamma_{j,k}(\alpha,\beta;x,y)\!=\\
 & \!\psi_{j,k}(\alpha,\beta;x,y)+(1-j)h(\alpha,\beta,1-\alpha-\beta),
\end{align*}
where ($x,y)$ is the unique positive solution of (\ref{eq:x0}) and
(\ref{eq:y0}). One can also show that the bound is exponentially
tight in $n$.
\end{IEEEproof}

\subsection{Scaling-Law Analysis for LM1 Stopping Sets}

For many CS problems, the primary interest is in scenarios where $\beta$
is small. This means that we need to perform stopping-set analysis
in the high-rate regime or to the signal vectors with sparse support.
For the convenience of analysis, we only derive the analysis for $(j,k)$
regular codes though it can be generalized to irregular codes \cite{orlitsky-it05}.
In our analysis, the variable node degree $j$ is fixed and the check
node degree $k$ is increasing. By calculating the scaling law of
$w_{j,k}(\beta)$, we find the uniform-in-probability recovery decoding
threshold $\beta_{j,k}^{*}$, which tells us the relationship between
the minimum number of measurements needed for uniform-in-probability
recovery and the sparsity of the signal. 

The following theorem shows the scaling law of LM1 for the $q$-SC.
\begin{thm}
\label{thm:LM1SS}There is a code from $(j,k)$ regular LDPC code
ensemble and a constant $K$ such that for the $q$-SC, all error
patterns of size $n\delta$ for $\delta<\bar{\beta}_{j}(k-1)^{-j/(j-2)}$
can be recovered by LM1 (w.h.p. as $n\rightarrow\infty$) for $k\ge K$
where $\bar{\beta}_{j}$ is the unique positive root in $c$ of the
following implicit function\vspace{-7mm} 
\begin{multline}
\\
v(d)=\frac{d}{2}\bigl((c-1)j\ln(1-c)-2c\ln(c)\\
+(1+c)(j-2)(-1+\ln d)\bigr)\\
\vspace{-10mm}\label{eq:v(d)}
\end{multline}
where $d=(1-c)^{-j/(j-2)}c^{2/(j-2)}$. 
\end{thm}

\begin{lem}
\label{lem:stp_lm1_1} Consider sequences of $(x_{k},y_{k})$ given
by (\ref{eq:x0}) and (\ref{eq:y0}), which satisfy $\beta_{k}=\Theta\left((k-1)^{-j/(j-2)}\right)$
as $k$ goes to infinity. In this case, the quantities $x_{k}$, $y_{k}$,
and $\alpha_{k}$ must all tend to zero.\end{lem}
\begin{IEEEproof}
See Appendix \ref{sec:Appendix5}.
\end{IEEEproof}

\begin{lem}
\label{lem:For-the--SC}For the $q$-SC with LM1 decoding and $j\geq3$,
the average number of stopping sets with size sublinear in $n$ goes
to zero as $n\rightarrow\infty$. More precisely, for each $3\leq j<k$
there exists a $\delta_{j,k}>0$ such that 
\[
\lim_{n\rightarrow\infty}\sum_{b=1}^{\delta_{j,k}n}\sum_{a=0}^{n-b}E_{n,j,k}\left(\frac{a}{n},\frac{b}{n}\right)=0.
\]
\end{lem}
\begin{IEEEproof}
See Appendix \ref{sec:Proof-of-Lemma7}.
\end{IEEEproof}

\begin{IEEEproof}
[Proof of Theorem \ref{thm:LM1SS}]The main idea of the proof is to
start from  (\ref{eq:gamma}) and find a scaling law for $w_{j,k}(\beta)$
as $k$ grows. Since $w_{j,k}(\beta)$ is the exponent of the average
number of stopping sets and the resulting scaling function $v(d)$
is negative in the range $\left(0,\bar{\beta}_{j}\right)$, almost
all codes have no stopping sets of size $n\delta$ with $0<\delta<\bar{\beta}_{j}(k-1)^{-j/(j-2)}$.
Because finding the limiting function of the scaled $w_{j,k}(\beta)$
is mathematically difficult, we first find an upper bound on $w_{j,k}(\beta)$
and then analyze the limiting function of this upper bound. 

Before we make any assumptions on the structure of $x$ and $y,$
we note that picking any $x$ and $y$ gives an upper bound of $\gamma_{j,k}(\alpha,\beta).$
To make the bound tight, we should pick good values for $x$ and $y.$
For example, the ($x,y$) that leads to the tightest bound is the
positive solution of  (\ref{eq:x0}) and  (\ref{eq:y0}). Since we
are free to choose the variables $x$ and $y$ arbitrarily, we assume
that $x$ and $y$ scale like $o\left(\frac{1}{k-1}\right)$. This
implies that the Taylor expansions of  (\ref{eq:x0}) and  (\ref{eq:y0})
converge. 

Applying Taylor expansion for small $x,y$ to (\ref{eq:x0}) and (\ref{eq:y0}),
we have
\begin{eqnarray*}
xy(k-1) & \approx & \alpha\\
(xy+y^{2}) & \approx & \beta.
\end{eqnarray*}
 Solving these equations for $x$ and $y$ gives the approximations
\begin{eqnarray*}
x_{0}\approx\frac{\alpha}{\sqrt{(\beta-\alpha)(k-1)}} &  & y_{0}\approx\sqrt{\frac{\beta-\alpha}{k-1}}.
\end{eqnarray*}
Next, we choose $\alpha=c\beta$ for $0<c<1$, which requires%
\footnote{The scaling regime we consider is $\beta=o(k^{-1})$ and this leads
to the scaling of $x,y$. This scaling of $x,y$ also implies that
$0<\alpha<\beta$. So we see that, although there exist stopping sets
with $\alpha\geq\beta$, they do not occur in the scaling regime we
consider. %
} that $0<\alpha<\beta$. Applying these substitutions to (\ref{eq:gamma})
gives 
\[
\gamma_{j,k}\left(c\beta,\beta;{\textstyle \frac{c\beta}{\sqrt{\beta(1-c)(k-1)}}},{\textstyle \sqrt{\frac{\beta(1-c)}{k-1}}}\right),
\]
which equals\vspace{-7mm}
\begin{multline}
\\
\frac{\beta}{2}\biggl(\left(1+c\right)\left(2-j\right)\left(1-\ln(\beta)\right)-\left(1-c\right)\, j\,\ln(1-c)\\
-2\, c\,\ln(c)+\left(1+c\right)\, j\,\ln(-1+k)\biggr)+O\left(\beta^{3/2}\right).\\
\vspace{-10mm}\label{eq:v(c,d)_small_beta}
\end{multline}
Plugging $\beta=d(k-1)^{-j/(j-2)}$ into this equation for $d\geq0$
gives\vspace{-7mm} 
\begin{multline}
\\
\gamma_{j,k}(\alpha,\beta)\le\frac{d}{2}(k-1)^{-j/(j-2)}\bigl((c-1)j\ln(1-c)-2c\ln(c)\\
+(1+c)(2-j)(1-\ln d)\bigr)+O\left((k-1)^{-2j/(j-2)}\right).\\
\vspace{-10mm}\label{eq:gamma_3}
\end{multline}

Scaling the RHS of  (\ref{eq:gamma_3}) by $(k-1)^{j/(j-2)}$ gives
the limiting function \vspace{-7mm}
\begin{multline}
\\
v(c,d)\triangleq\frac{d}{2}\bigl((c-1)j\ln(1-c)\\
-2c\ln(c)+(1+c)(2-j)(1-\ln d)\bigr).\\
\vspace{-10mm}\label{eq:v(c,d)}
\end{multline}
Next, we maximize the scaled upper bound of $\gamma_{j,k}(\alpha,\beta)$
over $\alpha$ by maximizing $v(c,d)$ over $c$. The resulting function
$v(d)\triangleq\max_{c\in(0,1)}v(c,d)$ is a scaled upper bound on
$w_{j,k}(\beta)$ as $k$ goes to infinity. Taking the derivative
w.r.t.\ $c$, setting it to zero, and solving for $d$ gives the
unique solution 
\begin{equation}
d=(1-c)^{-j/(j-2)}c^{2/(j-2)}.\label{eq:d_fun}
\end{equation}
Since the second derivative 
\[
\frac{d}{2}\left(-\frac{2}{c}-\frac{j}{1-c}\right)(k-1)^{-j/(j-2)}
\]
is negative, we have found a maximum. Moreover, $v(d)$ is given implicitly
by (\ref{eq:v(c,d)}) and (\ref{eq:d_fun}). The only positive root
of $v(d)$ is denoted $\bar{\beta}_{j}$ and is a constant independent
of $k.$ Fig. \ref{fig:v(d)} shows the curves given by numerical
evaluation of the scaled $w_{j,k}(\beta)$, which is given by 
\[
w'_{j,k}(d)=(k-1)^{j/(j-2)}w_{j,k}\left(d/(k-1)^{j/(j-2)}\right),
\]
and the limiting function $v(d).$ 
\begin{figure}
\centering{}\includegraphics[width=0.9\columnwidth]{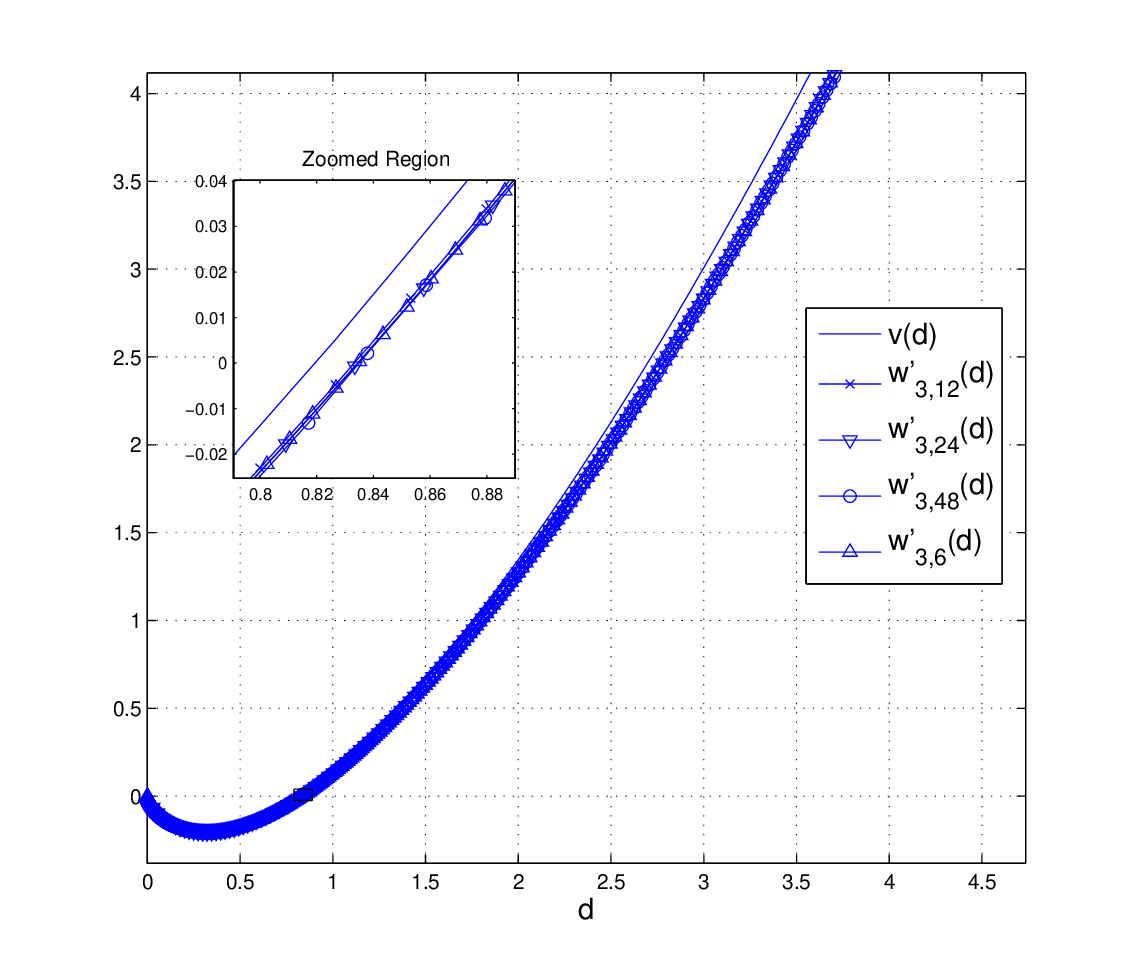}\caption{\label{fig:v(d)}Numerical evaluation $w'_{j,k}(d)$ and theoretical
bound $v(d)$}
\end{figure}

The proof is not yet complete, however, because we have not yet considered
stopping sets whose sizes are sublinear in $n$. To handle these,
we use Lemma \ref{lem:For-the--SC}, which shows that the average
number of stopping sets with size sublinear in $n$ also goes to zero.\end{IEEEproof}
\begin{rem}
In a CS system with strictly-sparse signals and LM1 reconstruction,
we have uniform-in-probability reconstruction (w.h.p. as $n\rightarrow\infty$)
of all signals with sparsity at most $n\delta$ where $\delta<\bar{\beta}_{j}(k-1)^{-j/(j-2)}$.
This requires $m=\gamma n\delta$ measurements and an oversampling
rate of $\gamma>\gamma_{0}=\bar{\beta}_{j}^{-(j-2)/j}j\delta^{-2/j}$.
\end{rem}

\begin{rem}
If the signal has all non-negative components, then the verification-based
algorithm will have no FV because the neighbors of a check node will
sum to zero only if these neighbors are exactly zero. Therefore, the
above analysis implies uniform recovery of non-negative signals that
are sufficiently sparse.
\end{rem}

\section{\label{sec:sparseCSit} Information Theory and Sparse CS}

As we mentioned in Section \ref{sec:Introduction}, many previous
works show that, for $p$-sparse signals of length-$n$, there is
a lower bound of $O\left(p\log(n/p)\right)$ on the number of measurements
for CS systems with noisy measurements \cite{Sarvotham-aller06,mlowerbound_crt-05,BaIPW10,BGIK-aller08}.
In general, these bounds can be obtained by thinking of the CS system
as a communication system and treating the measurements as different
observations of the sparse signal through the measurement channel.
The bound can be calculated by dividing the entropy in the unknown
sparse signal by the entropy obtained by each measurement. For the
cases that the entropy of the sparse signal scales as $O\left(p\log(n/p)\right)$
and the capacity of the measurement channel is finite, the lower bound
$O\left(p\log(n/p)\right)$ on the number of measurements is essentially
the \emph{best lower bound} shown in \cite{BaIPW10,BGIK-aller08}.
For example, let's consider a $p$-sparse signal with $1$'s at the
non-zero coefficients. The entropy of the signal is $\log\binom{n}{p}\geq p\log(n/p)$
bits. If the measurement is noisy, i.e., the capacity of the measurement
channel is finite, it is easy to see the minimum number of measurements
should scale as $O\left(p\log(n/p)\right)$ in order to recover the
signal. 

At first glance, the results in this paper seem to be at odds with
existing lower bounds on the number of measurements required for CS.
In this section, we explore the fundamental conditions for linear
scaling using sparse measurements from an information theoretic point
of view. 

Let $k$ and $j$ be check and variable degrees; let $n$ be the number
of variable nodes and $m=n^{\omega}$ be the number of check symbol
nodes. The random signal vector $X_{1}^{n}$ has i.i.d. components
drawn from $f_{X}(x)$ and the random measurement vector is $Y_{1}^{m}$.
The number of non-zero elements in the signal is controlled by assuming
that the average number of non-zero variable nodes attached to a check
node is given by $\lambda$. This allows us to write $f_{X}(x)=\frac{k-\lambda}{k}\delta(x)+\frac{\lambda}{k}f_{Z}(x)$,
where $Z$ is the random variable associated with a non-zero signal
element. Since $nj=mk$, the condition $\omega<1$ implies $k\rightarrow\infty$
and that the number of non-zero variable nodes attached to a check
node becomes Poisson with mean $\lambda$. Therefore, the amount of
information provided by the measurements is given by

\begin{align*}
H\left(Y_{1}^{m}\right) & \le\sum_{i=1}^{m}H\left(Y_{i}\right)\\
 & =\frac{nj}{k}\sum_{i=0}^{\infty}e^{-\lambda}\frac{\lambda^{i}}{i!}H(\underbrace{Z*Z*\cdots*Z}_{i\mbox{ times}})\\
 & \leq jn^{1-\omega}\sum_{i=0}^{\infty}e^{-\lambda}\frac{\lambda^{i}}{i!}(iH(Z))\\
 & =jn^{1-\omega}\lambda H(Z).
\end{align*}
Since $\lambda/k$ is the average fraction of non-zero variable nodes,
the entropy of the signal vector can be written as
\begin{align*}
H\left(X_{1}^{n}\right) & =-nh\left(\frac{\lambda}{k}\right)+n\frac{\lambda}{k}H(Z)\\
 & =\lambda n^{1-\omega}\ln\frac{1}{\lambda n^{-\omega}}+\lambda n^{1-\omega}H(Z)+O\left(n^{1-2\omega}\right).
\end{align*}
This implies that 
\[
H\!\left(Y_{1}^{m}\right)-H\!\left(X_{1}^{n}\right)\leq\lambda n^{1-\omega}\left((j-1)H(Z)-\; n\frac{1}{\lambda n^{-\omega}}\right).
\]
Since a necessary condition for reconstruction is $H\left(Y_{1}^{m}\right)-H\left(X_{1}^{n}\right)\geq0$,
we therefore find that 
\[
n\le\exp\left(\frac{H(Z)(j-1)+\ln\lambda}{\omega}\right)
\]
is required for reconstruction. This implies, that for any CS algorithm
to work, either $H(Z)$ has to be infinite or $j$ has to grow at
least logarithmically with $n.$ This does not conflict with the analysis
of LM2-MB for randomized reconstruction because, for signals over
real numbers or unbounded alphabets, the entropy $H(Z)$ can be infinite.

\section{\label{sec:Simulation-Results} Simulation Results}

\begin{figure}[t]
\centering{}\includegraphics[width=0.94\columnwidth]{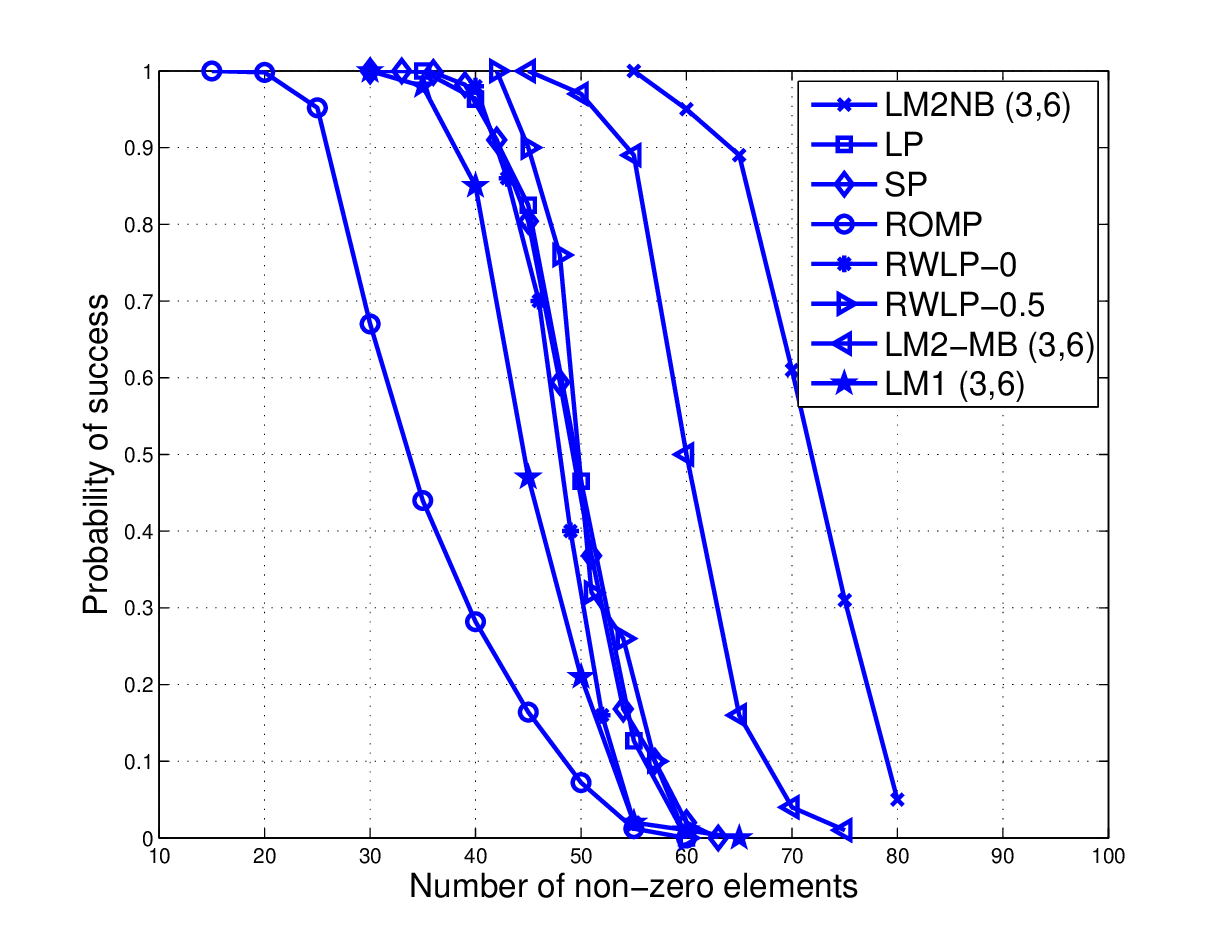}\caption{\label{fig:sim_zero-one} Simulation results for zero-one sparse signals
of length 256 with 128 measurements.}
\end{figure}

In this section, we provide the simulation results of LM1, LM2-MB
and LM2-NB reconstruction algorithms and compare these results with
other reconstruction algorithms. We consider two types of strictly-sparse
signals. The first type is the zero-one sparse signal where the entries
of the signal vector are either 0 or $\pm1$. The second type is the
Gaussian sparse case where the entries of the signal are either 0
or a Gaussian random variable with zero mean and unit variance. We
choose the signal length $n=256$ and number of measurements $m=128.$ 

\begin{figure}[t]
\centering{}\includegraphics[width=0.9\columnwidth]{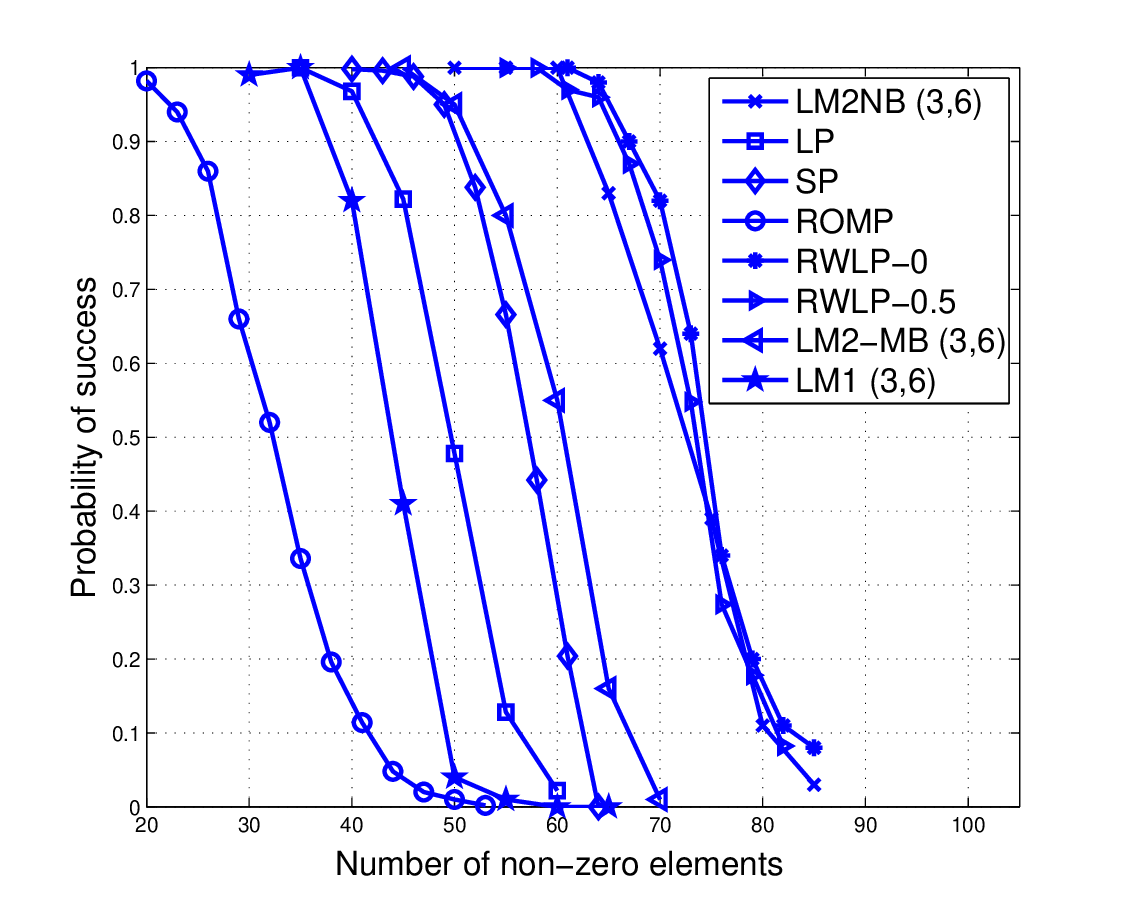}\caption{\label{fig:sim_Gaussian} Simulation results for Gaussian sparse signals
of length 256 with 128 measurements.}
\end{figure}

We compare different recovery algorithms such as linear-programming
(LP) \cite{Candes-it05}, subspace pursuit (SP) \cite{Dai-arxiv08},
regularized orthogonal matching pursuit (ROMP) \cite{Deanna-math09},
reweighted $\ell_{q}$ minimization (RWLP-$q$) \cite{chartrand_rwlp-08},
LM1, LM2-MB and LM2-NB. The measurement matrices for LM1, LM2-MB and
LM2-NB are generated randomly from the $(3,6)$, (4,8) and (5,10)
ensembles without double edges and 4-cycles. We also pick the non-zero
entries in the measurement matrices to be i.i.d. Gaussian random variables.
In all other algorithms, the measurement matrices are i.i.d. Gaussian
random matrices with zero mean and unit variance%
\footnote{We also tried the other algorithms with our sparse measurement matrices
(for the sake of fairness), but the performance was worse than the
dense Gaussian random matrices.%
}. Each point is obtained by simulating 100 blocks. Fig.~\ref{fig:sim_zero-one}
shows the simulation results for the zero-one sparse signal and Fig.~\ref{fig:sim_Gaussian}
shows the results for Gaussian sparse signal. From the results we
can see LM2-MB and LM2-NB perform favorably when compared to other
algorithms. 

\begin{figure}[b]
\centering{}\includegraphics[width=0.9\columnwidth]{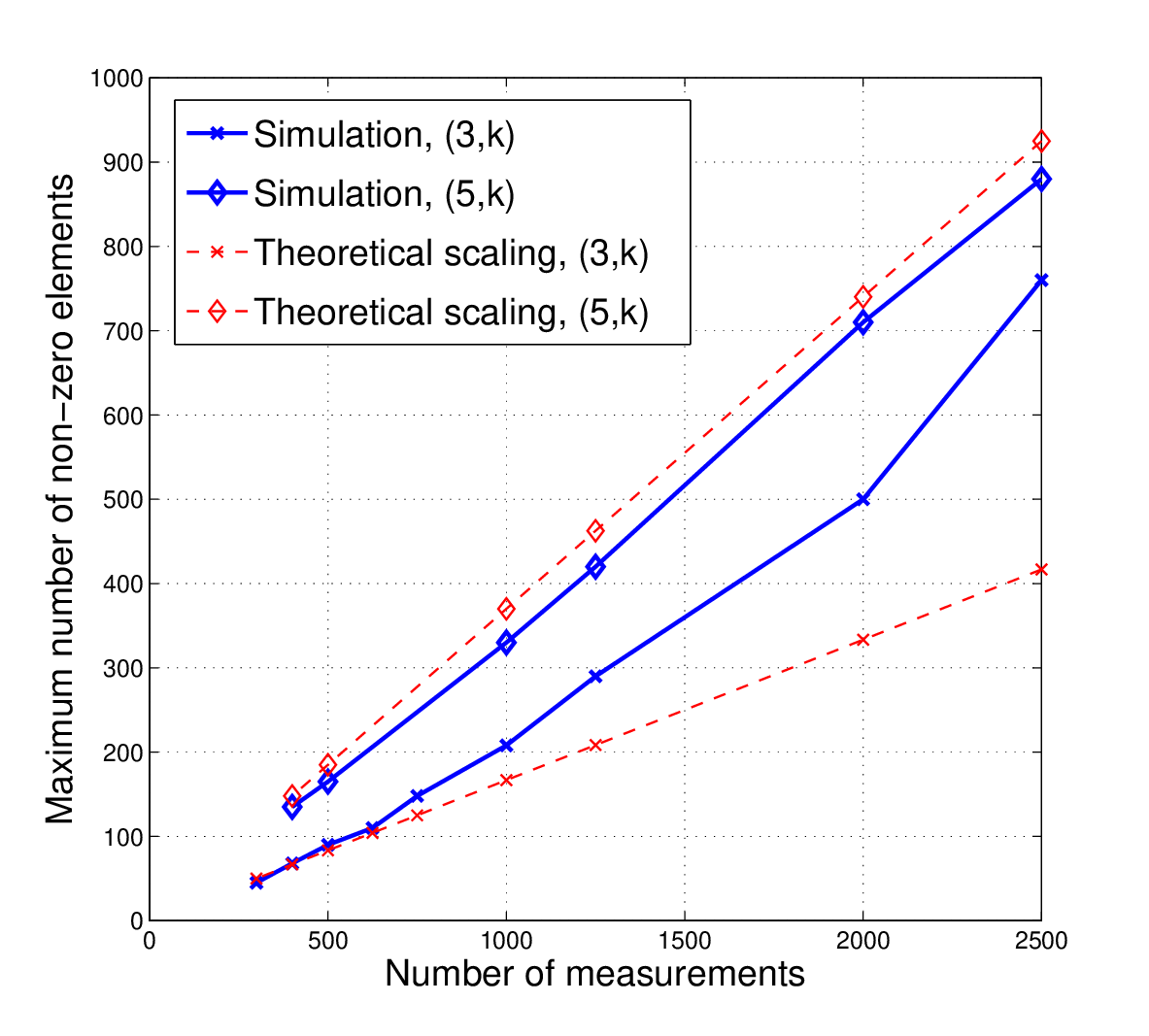}\caption{\label{fig:highratescaling} Simulation of high rate scaling of $(3,k)$
and $(5,k)$ ensembles for block length $n=10,000$.}
\end{figure}

From the simulation results, we can see LM2-NB outperforms LM2-MB.
In \cite{zp-it-lmp}, the authors provide details about the analysis
of LM2-NB and LM2-MB. In general, node-based algorithms perform better
than message-based algorithms for the same code. 

Another interesting observation is that LM1, LM2-MB and LM2-NB are
not sensitive to the magnitudes of the non-zero coefficients. They
perform almost the same for zero-one sparse signal and Gaussian sparse
signal. This is due to the verification-based nature of the decoding
algorithm. The other advantage of LM1 and LM2-NB is that they have
lower complexity in both the measuring process (i.e., encoding) and
the reconstruction process (i.e., decoding) than all other algorithms. 

In NB verification decoding, if the decoding finally succeeds, in
each iteration there is at least one node in the bipartite graph removed
due to verification. In each decoding iteration, all variable nodes
and check nodes are operating in parellel. Suppose that there is only
one node removed in each iteration, the number of multiplication and
addition operations on each node, or the time that a half-iteration
takes, is linearly with the check node degree $k$ (since we fix the
variable node degree $j$). Since the check node degree $k$ also
scales with $n$ and goes to infinity in our setting, the complexity
of a check-node operation is linearly with $k$. Notice that there
are $k$ variable nodes removed in each check node verification, the
complexity for removing each variable node is a constant independent
of $n$. Since there are $n$ variable nodes in the graph, the complexity
for successful decoding scales linearly with $n$. For LM1 algorithm,
it has equivalent MB and NB implementations \cite{zp-it-lmp}. Therefore,
the complexity of LM1 also scales linearly with $n$. 

For LM2-MB algorithm, if the variable and check node degrees are constants
independent of $n$, it is easy to show the linearity of complexity.
However, in our setting, check node degree $k$ goes to infinity as
$n$ goes to infinity. We cannot show the complexity is linearly in
$n$. Fortunately, for small $j$ and $k$, and large $n$, LM2-MB
runs almost as fast as NB algorithm based on our simulation.

We also find the maximum sparsity $p^{*}$ for perfect reconstruction
when we use parity-check matrices from $(3,k)$ and $(5,k)$ ensembles
(with different $k$) as the measurement matrices when $n$ is large
and try to see how $p^{*}$ scales with the code rate. In the simulation,
we fix $n=10000$, try different $k$'s (or $m$'s) and use LM2-MB
as the decoding algorithm. Fig.~\ref{fig:highratescaling} shows
the how $p^{*}$ scales with $m$ in high-rate regime. We also show
the theoretical scaling in Fig.~\ref{fig:highratescaling}, which
is $\bar{\alpha}_{j}nj/k$ with $\bar{\alpha}_{3}=1/6$ and $0.3723<\bar{\alpha}_{5}<0.3724$.
Since we are considering high-rate scaling as $k\rightarrow\infty$
and fixing $j$, which also means $\frac{m}{n}=\frac{j}{k}\rightarrow0$,
where $m$ is the number of measurements. Therefore, our results are
more accurate when $m$ is small. Notice that the simulation and the
theoretical results match very well in the high-rate region. 

\begin{figure}[t]
\centering{}\includegraphics[width=0.95\columnwidth]{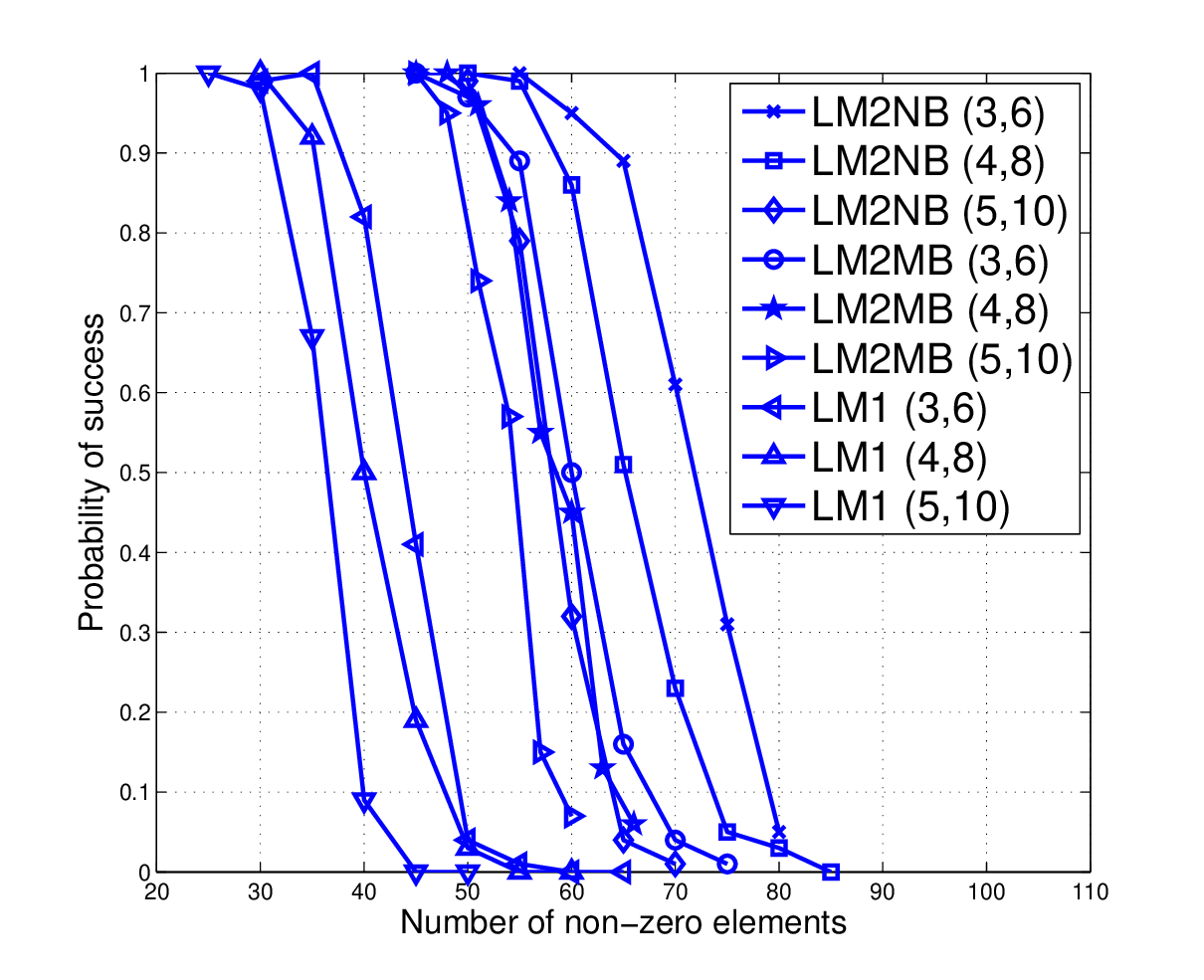}\caption{\label{fig:sim_zero-one_345} Simulation results for zero-one spikes
of length 256 with 128 measurements by using $(3,6)$, $(4,8)$ and
$(5,10)$ ensembles.}
\end{figure}

\begin{figure}[t]
\centering{}\includegraphics[width=0.9\columnwidth]{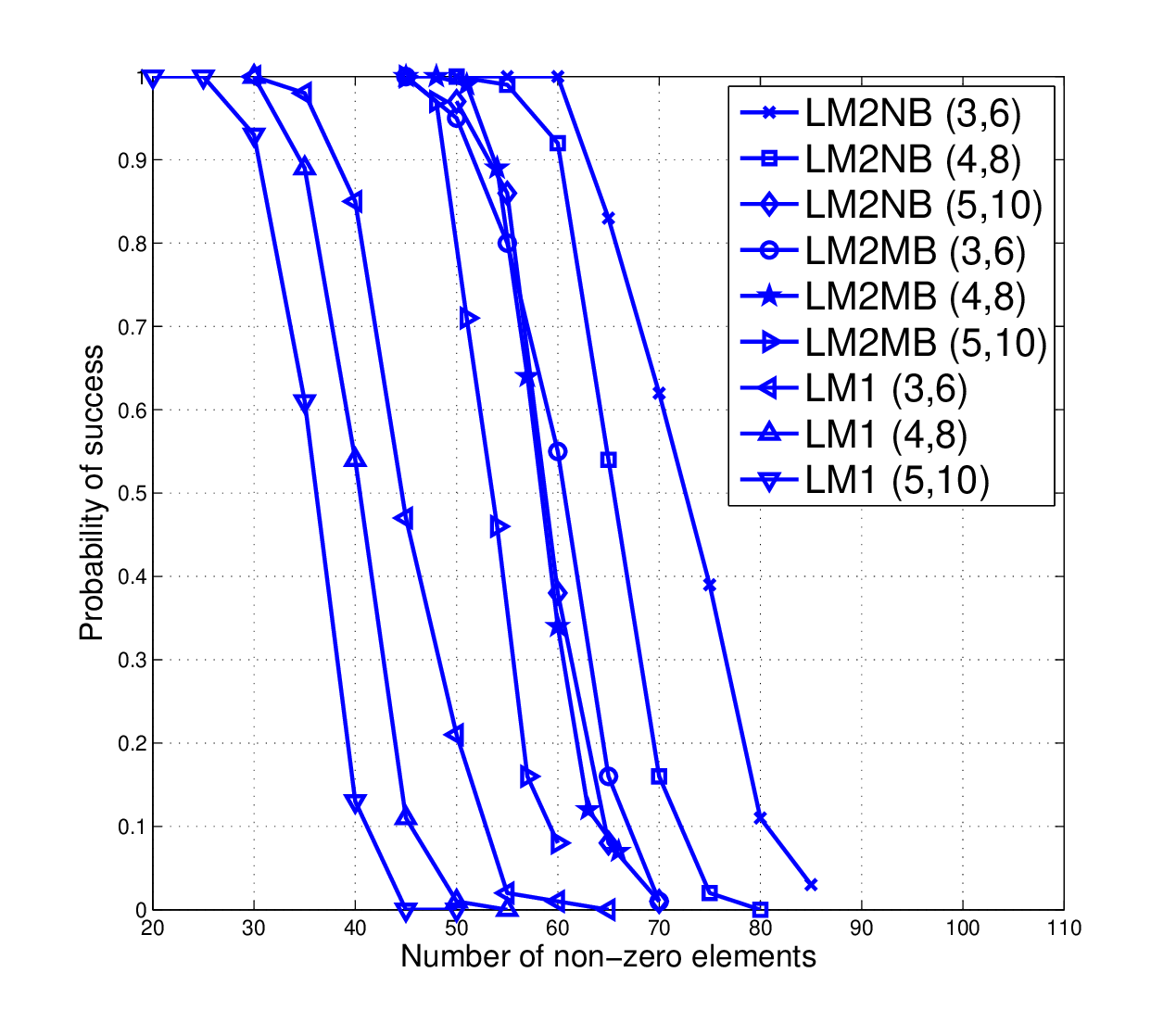}\caption{\label{fig:sim_Gaussian_345} Simulation results for Gaussian spikes
of length 256 with 128 measurements by using $(3,6)$, $(4,8)$ and
$(5,10)$ ensembles.}
\end{figure}

The simulation results for $(3,6)$, $(4,8)$ and $(5,10)$ ensembles
are shown in Fig.~\ref{fig:sim_zero-one_345} and Fig.~\ref{fig:sim_Gaussian_345}.
The results show that for short block length and rate a half, using
measurement matrix from ensemble with higher VN/CN degree leads to
worse performance. This seems to conflict the results shown in Fig.~\ref{fig:highratescaling},
since the results in Fig.~\ref{fig:highratescaling} show that (5,$k$)
ensemble should perform better than (3,$k$) ensemble. The reason
for this is that our scaling-law analysis is only accurate when code
rate is high. In the scaling-law analysis, we consider rates close
to 1 and large block-length, which is not satisfied in the simulation
of Fig.~\ref{fig:sim_zero-one_345} and Fig.~\ref{fig:sim_Gaussian_345}.

\section{Conclusion\label{sec:Conclusion}}

We analyze message-passing decoding algorithms for LDPC codes in the
high-rate regime. The results can be applied to compressed sensing
systems with strictly-sparse signals. A high-rate analysis based on
DE is used to derive the scaling law for randomized reconstruction
CS systems and stopping-set analysis is used to analyze uniform-in-probability/uniform
reconstruction. The scaling-law analysis gives the surprising result
that LDPC codes, together with the LM2-MB algorithm, allow randomized
reconstruction when the number of measurements scales linearly with
the sparsity of the signal. Simulation results and comparisons with
a number of other CS reconstruction algorithms are also provided. 

\appendices

\section{Proof of Proposition \ref{cor:BEC-alpha-thr} \label{sec:Appendix1}}
\begin{IEEEproof}
Starting with the convergence condition $\lambda\left(1-e^{-\overline{\alpha}_{j}jx}\right)\leq x$
for $x\in(0,1]$, we first solve for $\overline{\alpha}_{j}$ to get
\begin{equation}
\overline{\alpha}_{j}=\inf_{x\in(0,1)}-\frac{1}{jx}\ln\left(1-\lambda^{-1}(x)\right).
\end{equation}
Next, we substitute $x=\lambda\left(1-e^{-y}\right)$ and simplify
to get
\begin{equation}
\overline{\alpha}_{j}=\inf_{y\in(0,\infty)}\frac{y}{j\lambda\left(1-e^{-y}\right)}.
\end{equation}
For $j\geq3$, this function is unbounded as $y\rightarrow0$ or $y\rightarrow\infty$,
so the minimum must occur at an interior critical point $y^{*}$.
Choosing $\lambda(x)=x^{j-1}$ and setting the derivative w.r.t. $y$
to zero gives 
\begin{equation}
\frac{j\left(1-e^{-y^{*}}\right)^{j-1}-j(j-1)y^{*}\left(1-e^{-y^{*}}\right)^{j-2}e^{-y^{*}}}{j^{2}\left(1-e^{-y^{*}}\right)^{2j-2}}=0
\end{equation}
Canceling terms and simplifying the numerator gives $1-e^{-y^{*}}-(j-1)y^{*}e^{-y^{*}}=0$,
which can be rewritten as $e^{y^{*}}=(j-1)y^{*}+1$. Ignoring $y^{*}=0$,
this implies that $y^{*}$ is given by the unique intersection of
$e^{y}$ and $(j-1)y+1$ for $y>0$. That intersection point can be
written in closed form using the non-principal real branch of the
Lambert W-function \cite{Corless-aicm96}, $W_{-1}(x)$, and is given
by, for $j\geq2$, 
\begin{equation}
y_{j}^{*}=-\frac{1}{j-1}\left(1+(j-1)W_{-1}\left(-\frac{1}{j-1}e^{-1/(j-1)}\right)\right).
\end{equation}
Using this, the $\alpha$-threshold for $j$-regular ensembles is
given by $\overline{\alpha}_{j}=\frac{1}{j}y_{j}^{*}\left(1-e^{-y_{j}^{*}}\right)^{1-j}$.
For $j=2$, the minimum occurs as $y_{2}^{*}\rightarrow0$ and the
limit gives $\overline{\alpha}_{2}=\frac{1}{2}$.
\end{IEEEproof}

\section{Proof of Lemma \ref{lem:Let-the-functions} \label{sec:Appendix2}}
\begin{IEEEproof}
All statements are implied to hold for all $k>\overline{\alpha}_{j}^{j-1}$,
all $x\in[0,1]$, and all $\alpha\in[0,\overline{\alpha}_{j}]$. Since
$1-(1-x)^{k}$ is concave for $k\geq1$, the tangent upper bound at
$x=0$ shows that $1-(1-x)^{k}\leq kx$. This implies that
\begin{equation}
\left(1-\left(1-\frac{\overline{\alpha}_{j}x}{k^{j/(j-1)}}\right)^{k}\right)^{j-1}\leq\frac{\overline{\alpha}_{j}^{j-1}x^{j-1}}{k}.\label{eq:LM1DE_Lemma_Eqn1}
\end{equation}
Since $\frac{\alpha}{k^{j/(j-1)}}\leq\frac{\overline{\alpha}_{j}}{\overline{\alpha}_{j}^{j-1}}\leq1$,
we can use (\ref{eq:LM1DE_Lemma_Eqn1}) to upper bound $g_{k+1}(x)$
with\vspace{-7mm}

\begin{multline*}
\\
g_{k+1}(x)\le\frac{\alpha}{\overline{\alpha}_{j}}\Biggl(1-\Biggl[1-\frac{\overline{\alpha}_{j}^{j-1}x^{j-1}}{k}\\
+\frac{\alpha}{k^{j/(j-1)}}\frac{\overline{\alpha}_{j}^{j-1}x^{j-1}}{k}-\frac{\overline{\alpha}_{j}x}{k^{j/(j-1)}}\Biggr]^{k}\Biggr)^{j-1}\\
\le\frac{\alpha}{\overline{\alpha}_{j}}\Biggl(1-\Biggl[1-\frac{\overline{\alpha}_{j}^{j-1}x^{j-1}}{k}-\frac{\overline{\alpha}_{j}x}{k^{j/(j-1)}}\Biggr]^{k}\Biggr)^{j-1}.\\
\vspace{-10mm}
\end{multline*}
 This completes the proof of (i). 

The fact that $\overline{g}_{k+1}(x)$ is monotonically decreasing
follows from Lemma \ref{lem:MonotonicToExp}. This completes the proof
of (ii). Lemma \ref{lem:MonotonicToExp} also shows that the limit
of $\overline{g}_{k+1}(x)$ is 
\[
g_{*}(x)\triangleq\frac{\alpha}{\overline{\alpha}_{j}}\left(1-e^{-\overline{\alpha}_{j}^{j-1}x^{j-1}}\right)^{j-1}.
\]
This proves the first part of (iii).

Next, we will show that 
\[
\lim_{k\rightarrow\infty}g_{k}(x)=\frac{\alpha}{\overline{\alpha}_{j}}\left(1-e^{-\overline{\alpha}_{j}^{j-1}x^{j-1}}\right)^{j-1}.
\]
 First, we show that
\begin{equation}
\lim_{k\rightarrow\infty}k\left(1-\left(1-\frac{\overline{\alpha}_{j}x}{k^{j/(j-1)}}\right)^{k}\right)^{j-1}\!\!\!\!\!\!=\overline{\alpha}_{j}^{j-1}x^{j-1}.\label{eq:LM1DE_limit1}
\end{equation}
In light of the the upper bound (\ref{eq:LM1DE_Lemma_Eqn1}), the
limit is clearly upper bounded by $\overline{\alpha}_{j}^{j-1}x^{j-1}$.
Using the lower bound in Lemma \ref{lem:MonotonicToExp}, we see that
\begin{align*}
\left(1-\frac{\overline{\alpha}_{j}x}{k^{j/(j-1)}}\right)^{k} & \geq\frac{e^{-\overline{\alpha}_{j}xk^{-1/(j-1)}}}{\left(1+\overline{\alpha}_{j}xk^{-j/(j-1)}\right)^{\overline{\alpha}_{j}xk^{-1/(j-1)}}}\\
 & \geq\frac{e^{-\overline{\alpha}_{j}xk^{-1/(j-1)}}}{\left(1+\overline{\alpha}_{j}xk^{-j/(j-1)}\right)}\\
 & \geq\left(1-\frac{\overline{\alpha}_{j}x}{k^{j/(j-1)}}\right)e^{-\overline{\alpha}_{j}xk^{-1/(j-1)}}.
\end{align*}
This implies that {\footnotesize 
\[
\left(\!1\!-\!\left(\!1\!-\!\frac{\overline{\alpha}_{j}x}{k^{j/(j-1)}}\!\right)^{k}\!\right)^{j-1}\!\!\!\!\!\!\!\!\ge\left(\!1\!-\!\left(\!1\!-\!\frac{\overline{\alpha}_{j}x}{k^{j/(j-1)}}\!\right)e^{-\overline{\alpha}_{j}xk^{-1/(j-1)}}\right)^{j-1}.
\]
}Together with{\small 
\[
\lim_{k\rightarrow\infty}k\left(1-\left(1-\frac{\overline{\alpha}_{j}x}{k^{j/(j-1)}}\right)e^{-\overline{\alpha}_{j}xk^{-1/(j-1)}}\right)^{j-1}=\overline{\alpha}_{j}^{j-1}x^{j-1},
\]
}we see that the limit (\ref{eq:LM1DE_limit1}) holds.

To calculate the limit of $g_{k}(x)$, we can use the fact that $\lim_{k\rightarrow\infty}\left(1-a_{k}+o\left(\frac{1}{k}\right)\right)^{k}=e^{-\lim_{k\rightarrow\infty}ka_{k}}$
whenever $\lim_{k\rightarrow\infty}ka_{k}$ exists. Using this, we
see that $\lim_{k\rightarrow\infty}g_{k+1}(x)$ can be rewritten as\vspace{-7mm}
{\footnotesize 
\begin{multline*}
\\
\lim_{k\rightarrow\infty}\frac{\alpha}{\overline{\alpha}_{j}}\left(1\!-\left[1\!-\left(1\!-\left(1\!-\frac{\overline{\alpha}_{j}x}{k^{j/(j-1)}}\right)^{k}\right)^{j-1}\!\!\!\!\!\!\!\!\!+o\left(\frac{1}{k}\right)\right]^{k}\right)^{j-1}\\
=\frac{\alpha}{\overline{\alpha}_{j}}\left(1-e^{-\overline{\alpha}_{j}^{j-1}x^{j-1}}\right)^{j-1},\\
\vspace{-10mm}
\end{multline*}
}where the last step follows from (\ref{eq:LM1DE_limit1}). 
\end{IEEEproof}

\section{Proof of Corollary \ref{lem:lm1_alpha} \label{sec:Appendix3}}
\begin{IEEEproof}
Recall that $\bar{\alpha}_{j}$ is defined as the largest $\alpha$
s.t. $\left(1-e^{-\alpha^{j-1}x^{j-1}}\right)^{j-1}<x$ for $x\in(0,1]$.
Solving this inequality for $\alpha$ allows one to express $\bar{\alpha}_{j}$
as 
\begin{equation}
\bar{\alpha}_{j}=\inf_{x\in(0,1]}h_{j}(x)\label{eq:alpha_bar_lm1}
\end{equation}
where $h_{j}(x)=\left(-\ln\left(1-x^{1/(j-1)}\right)x^{(1-j)}\right)^{1/(j-1)}.$
Since $-\ln\left(1-x^{1/(j-1)}\right)\geq x^{1/(j-1)}$, it follows
that $h_{j}(x)\geq x^{1/(j-1)^{2}-1}\geq1$ for $x\in(0,1]$. Therefore,
$\bar{\alpha}_{j}\geq1$ for $j\geq2$. 

Notice that $h_{j}(x)$ is a monotonically increasing function of
$x$ when $j=2.$ So we have 
\begin{equation}
\bar{\alpha}_{2}=\lim_{x\rightarrow0}h_{j}(x)=1.
\end{equation}
When $j\ge3,$ $h_{j}(x)$ goes to infinity when $x$ goes to either
0 or 1, so the infimum is achieved at an interior point $x_{j}^{*}$.
By taking derivative of $x$ and setting it to zero, $x_{j}^{*}$
is the solution of 
\begin{equation}
\frac{x^{\frac{1}{j-1}}}{\left(1-x^{\frac{1}{j-1}}\right)\ln\left(1-x^{\frac{1}{j-1}}\right)}=-\left(j-1\right)^{2}.
\end{equation}
 So 
\begin{equation}
x_{j}^{*}=\left(1+\frac{1}{(j-1)^{2}W_{-1}\left(-e^{-1/(j-1)^{2}}/(j-1)^{2}\right)}\right)^{2}.\label{eq:eq12}
\end{equation}
By solving this numerically, we find that $x_{3}^{*}=0.816042,$ $x_{4}^{*}=0.938976$
and $x_{5}^{*}=0.971087.$ Substituting $x_{j}^{*}$ into  (\ref{eq:alpha_bar_lm1}),
we have $1.87321<\bar{\alpha}_{3}<1.87322,$ $1.66455<\bar{\alpha}_{4}<1.66456$
and $1.52073<\bar{\alpha}_{5}<1.52074$.
\end{IEEEproof}

\section{Proof of Lemma \ref{lem:LM2_bound} \label{sec:Appendix4}}
\begin{IEEEproof}
Let us define the function $\hat{g}_{k}(x)$ with\vspace{-7mm}

\begin{multline*}
\\
\hat{g}_{k}(x)\triangleq\frac{\alpha}{\bar{\alpha}_{j}}\left(1-\left(1-\frac{\alpha jx}{k}\right)^{k-1}\right)^{j-1}\\
+\left(j-1\right)\left(1-\left(1-\frac{\alpha jx}{k}\right)^{k-1}\right)^{j-2}\left(1-\frac{\alpha jx}{k}\right)^{k-1}.\\
\vspace{-10mm}
\end{multline*}
To prove (i), we will show $g_{k}(x)<\hat{g}_{k}(x)<\bar{g}_{k}(x).$
To see that $g_{k}(x)<\hat{g}_{k}(x),$ we must simply observe that
\[
1-\left(1-\frac{1-\frac{\alpha j}{k}}{1-\frac{\alpha jx}{k}}\left(1-\left(1-\frac{\alpha jx}{k}\right)^{k-1}\right)^{j-1}\right)^{k-1}<1.
\]
This can be seen by working from the inner expression outwards and
using the facts that $0<\frac{\alpha j}{k}<1$ and $0<x<1$. Each
step gives a result that is bounded between 0 and 1.

To show $\hat{g}_{k}(x)<\bar{g}_{k}(x),$ we first change variables
to $z=\left(1-\frac{\alpha jx}{k}\right)^{k}$ where $z\in(0,1)$.
This allows $\bar{g}_{k}(x)$ to be written as a function of $z$
with
\begin{equation}
\bar{g}_{k}(z)=\frac{\alpha}{\bar{\alpha}_{j}}\left(\left(1-z\right)^{j-1}+\left(j-1\right)\left(1-z\right)^{j-3}z\right).
\end{equation}
Taking the derivative of $\bar{g}_{k}(z)$ with respect to $z$ gives
\begin{equation}
\frac{\mbox{d}\bar{g}_{k}(z)}{\mbox{d}z}=-\frac{\alpha}{\bar{\alpha}_{j}}\left(j-2\right)\left(j-1\right)\left(1-z\right)^{j-3}z,
\end{equation}
which is negative for $j\geq3$. So $\bar{g}_{k}(z)$ is a monotonically
decreasing function of $z.$ Using the inequality $\left(1-\frac{\alpha jx}{k}\right)^{k-1}>\left(1-\frac{\alpha jx}{k}\right)^{k},$
we find that $\hat{g}_{k}(x)<\bar{g}_{k}(x)$. 

Next, we will prove (ii) by showing the limits of $g_{k}(x)$ and
$\bar{g}_{k}(x)$ are the same. First, we take the the term by term
limit of $\bar{g}_{k}(x)$ to see that

\begin{align}
\lim_{k\rightarrow\infty}\bar{g}_{k}(x)= & \frac{\alpha}{\bar{\alpha}_{j}}\Bigl(\left(1-e^{-\alpha jx}\right)^{j-1}+\nonumber \\
 & \;\;(j-1)\left(1-e^{-\alpha jx}\right)^{j-2}e^{-\alpha jx}\Bigr)\nonumber \\
= & \frac{\alpha}{\bar{\alpha}_{j}}\left(1-e^{-\alpha jx}\right)^{j-2}\left(1+(j-2)e^{-\alpha jx}\right).\label{eq:lm2_gkbar_limit}
\end{align}
Next, we use the fact that $\left(1-\frac{\alpha jx}{k}\right)^{k-1}\rightarrow e^{-\alpha jx}$
to see that
\[
\lim_{k\rightarrow\infty}\left(1-\frac{1-\frac{\alpha j}{k}}{1-\frac{\alpha jx}{k}}\left(1-\left(1-\frac{\alpha jx}{k}\right)^{k-1}\right)^{j-1}\right)^{k-1}=0.
\]
From this, we find that the term by term limit of $g_{k}(x)$ is also
equal to (\ref{eq:lm2_gkbar_limit}).

To prove (iii), we recall that, using the change of variables $z=\left(1-\frac{\alpha jx}{k}\right)^{k}$,
$\bar{g}_{k}(z)$ is a monotonically decreasing function of $z$.
Moreover, $\bar{g}_{k}(z)$ does not depend on $k$ and $z=\left(1-\frac{\alpha jx}{k}\right)^{k}$
is a monotonically increasing function of $k$ (e.g., see Lemma \ref{lem:MonotonicToExp}).
So $\bar{g}_{k}(x)$ is a monotonically decreasing function of $k.$
\end{IEEEproof}

\section{Proof of Lemma \ref{lem:stp_lm1_1} \label{sec:Appendix5}}
\begin{IEEEproof}
Consider whether the sequences $x_{k}$ and $y_{k}$ converge to zero
or not. Clearly, there are only 4 possible cases.

If $x_{k}=o(1)$ and $y_{k}=\Omega(1)$, the limit
\begin{equation}
\lim_{k\rightarrow\infty}\beta_{k}=\lim_{k\rightarrow\infty}\frac{y_{k}(1+y_{k})^{k-1}}{(1+y_{k})^{k}}=\lim_{k\rightarrow\infty}y_{k}
\end{equation}
 contradicts $\beta_{k}=\Theta\left((k-1)^{-j/(j-2)}\right).$ 

If $x_{k}=\Omega(1)$ and $y_{k}=o(1)$, the limit 
\[
\lim_{k\rightarrow\infty}k\beta_{k}=\lim_{k\rightarrow\infty}\frac{y_{k}\left(1+x_{k}\right)^{k-1}}{y_{k}\left(1+x_{k}\right)^{k-1}}=1
\]
 contradicts $\beta_{k}=\Theta\left((k-1)^{-j/(j-2)}\right)$. 

If $x_{k}=\Omega(1)$ and $y_{k}=\Omega(1)$, the limit satisfies
\begin{align*}
\lim_{k\rightarrow\infty}\beta_{k} & =\lim_{k\rightarrow\infty}\frac{y_{k}\left(1+x_{k}+y_{k}\right)^{k-1}}{\left(1+x_{k}+y_{k}\right)^{k}-\left(1+x_{k}\right)^{k}}\\
 & >\lim_{k\rightarrow\infty}\frac{y_{k}}{1+x_{k}+y_{k}},
\end{align*}
and this contradicts $\beta_{k}=\Theta\left((k-1)^{-j/(j-2)}\right)$.
\end{IEEEproof}

\section{Proof of Lemma\label{sec:Proof-of-Lemma7} \ref{lem:For-the--SC}}

Since all stopping sets with size sublinear in $n$ shrink to the
zero point on the scaled curve, we must treat sublinear stopping sets
separately. The proof proceeds by considering separately stopping
sets of size $O(\ln n)$ and size $\delta n$ for very small $\delta$.
The number of correct and incorrect variable nodes in a stopping set
is denoted, respectively, $a$ and $b$ (i.e., $n\alpha=a$ and $n\beta=b$).
\begin{IEEEproof}
Using (\ref{eq:average_number_ss}) and Lemma \ref{lem:bounds_on_multinomial},
we can bound $E_{n,j,k}(\alpha,\beta)$ with 
\[
E_{n,j,k}(\alpha,\beta)\!\le\! je^{\frac{1}{12jn}}e^{(1-j)nh(\alpha,\beta,1-\alpha-\beta)}S_{n,j,k}(\alpha n,\beta n).
\]
The coefficient $S_{n,j,k}(a,b)$ can be bounded using a Chernoff-type
bound and this gives\vspace{-7mm}
\begin{multline*}
\\
\ln S_{n,j,k}(a,b)\leq\frac{jn}{k}\ln\frac{1\!+\!(1\!+\! x\!+\! y)^{k}\!-\! ky\!-\!(1\!+\! x)^{k}}{x^{ja}y^{jb}}\\
\leq\frac{jn}{k}\ln\left((1\!+\! x\!+\! y)^{k}\!-\! ky-\! kx\right)\!-\! ja\ln x\!-\! jb\ln y\\
\vspace{-10mm}
\end{multline*}
for arbitrary $x\geq0$ and $y\geq0$. Choosing $x=\frac{1}{\sqrt{n}}$
and $y=\frac{1}{\sqrt{n}}$ gives the bound
\begin{align}
S_{n,j,k}(a,b) & \le e^{2j(k-1)+O(n^{-1/2})}n^{(a+b)j/2}\nonumber \\
 & \leq Cn^{(a+b)j/2},\label{eq:bnd_A}
\end{align}
where $C$ is a constant independent of $n$. Applying  (\ref{eq:bnd_A})
to the $E_{n,j,k}(\alpha,\beta)$ bound shows that 
\begin{align}
E_{n,j,k}\left(\frac{a}{n},\frac{b}{n}\right) & \leq je^{\frac{1}{12nj}}\exp\left((1-j)nh\left(\frac{a}{n},\frac{b}{n},1-\frac{a}{n}-\frac{b}{n}\right)\right)S_{n,j,k}(a,b)\nonumber \\
 & \leq je^{\frac{1}{12nj}}\left(\frac{a}{n}\right)^{(j-1)a}\left(\frac{b}{n}\right)^{(j\!-\!1)b}S_{n,j,k}(a,b)\nonumber \\
 & \leq je^{\frac{1}{12j}}Cn^{(a+b)(j/2-(j-1)(1-\epsilon))}\left(\frac{a}{n^{\epsilon}}\right)^{(j-1)a}\left(\frac{b}{n^{\epsilon}}\right)^{(j-1)b}\label{eq:Enjk_Simple_Chernoff}
\end{align}
 where $0<\epsilon<\frac{1}{4}$ and $j\ge3$. 

Now, we can use this to show that
\[
\lim_{n\rightarrow\infty}\sum_{b=1}^{A\ln n}\sum_{a=0}^{n-b}E_{n,j,k}\left(\frac{a}{n},\frac{b}{n}\right)=0.
\]
Since a stopping set cannot have a check node that attaches to only
verified and correct edges, a simple counting argument shows that
$S_{n,j,k}(a,b)=0$ if $a>(k-1)b$. Therefore, the above condition
can be simplified to
\begin{equation}
\lim_{n\rightarrow\infty}\sum_{b=1}^{A\ln n}\sum_{a=0}^{(k-1)b}E_{n,j,k}\left(\frac{a}{n},\frac{b}{n}\right)=0.\label{eq:double_sum}
\end{equation}
Starting from (\ref{eq:Enjk_Simple_Chernoff}), we note that $b\leq A\ln n$
and $a\le(k-1)b$ implies that $\left(\frac{a}{n^{\epsilon}}\right)^{(j-1)a}\left(\frac{b}{n^{\epsilon}}\right)^{(j-1)b}<1$
for large enough $n$. Therefore, we find that the double sum in (\ref{eq:double_sum})
is upper bounded by
\[
je^{\frac{1}{12j}}Cn^{(j/2-(j-1)(1-\epsilon))}(k-1)(A\ln n)^{2}
\]
for large enough $n$. Since the exponent $(a+b)(j/2-(j-1)(1-\epsilon))$
of $n$ is negative as long as $\epsilon<\frac{1}{4}$ and $j\geq3$,
we also find that the limit of the double sum in (\ref{eq:double_sum})
goes to zero as $n$ goes to infinity for any $A>0$. 

Now, we consider stopping sets of size greater than $A\ln n$ but
less than $\delta_{j,k}n$. Combining (\ref{eq:gamma}) and Lemma
\ref{lem:bounds_on_multinomial} shows that $E_{n,j,k}(\alpha,\beta)\le je^{\frac{1}{12jn}}e^{n\gamma_{j,k}(\alpha,\beta)}.$
Notice that (\ref{eq:v(c,d)_small_beta}) is an accurate upper bound
on $\gamma_{j,k}(\alpha,\beta)$ for small enough $\beta$ and its
maximum over $\alpha$ is given parametrically by (\ref{eq:v(d)}).
Moreover, $v(d)$ is strictly decreasing at $d=0$, and this implies
that $\gamma_{j,k}(\alpha,\beta)$ is strictly decreasing in $\beta$
at $\beta=0$ for all valid $\alpha$. Therefore, there is a $\delta_{j,k}>0$
and $\eta>0$ such that
\[
\gamma_{j,k}(\alpha,\beta)<-\eta\beta
\]
for all $0\leq\beta\leq\delta_{j,k}$. From this, we conclude that
$\frac{A\ln n}{n}<\beta<\delta_{j,k}n$, which implies that\vspace{-7mm}
\begin{multline*}
\\
E_{n,j,k}(\alpha,\beta)\le je^{\frac{1}{12jn}}e^{n\gamma_{j,k}(\alpha,\beta)}\leq je^{\frac{1}{12jn}}e^{-n\eta\beta}\\
\leq je^{\frac{1}{12jn}}e^{-\eta A\ln n}\le je^{\frac{1}{12jn}}n^{-A\eta},\\
\vspace{-10mm}
\end{multline*}
where $A\eta$ can be made arbitrarily large by increasing $A$. Choosing
$A=\frac{3}{\eta}$ so that $A\eta=3$ shows that
\[
\lim_{n\rightarrow\infty}\sum_{b=\log n}^{\delta_{j,k}n}\sum_{a=0}^{n-b}E_{n,j,k}\left(\frac{a}{n},\frac{b}{n}\right)\leq\lim_{n\rightarrow\infty}n^{2}je^{\frac{1}{12jn}}n^{-3}=0.
\]
 This completes the proof.
\end{IEEEproof}

\section{Lemma \ref{lem:bounds_on_multinomial} \label{sec:Appendix6}}
\begin{lem}
\label{lem:bounds_on_multinomial} The ratio $D\triangleq\frac{\left(\begin{array}{c}
n\\
a,b,n-a-b
\end{array}\right)}{\left(\begin{array}{c}
nj\\
aj,bj,(n-a-b)j
\end{array}\right)}$ can be bounded with\vspace{-7mm}
\begin{multline*}
\\
j\exp\left((1-j)nh(\frac{a}{n},\frac{b}{n},1-\frac{a}{n}-\frac{b}{n})-\frac{1}{12n}\right)\le D\\
\le j\exp\left((1-j)nh(\frac{a}{n},\frac{b}{n},1-\frac{a}{n}-\frac{b}{n})+\frac{1}{12jn}\right).\\
\vspace{-10mm}
\end{multline*}
\end{lem}
\begin{IEEEproof}
Let $D$ be defined by
\[
D=\frac{\left(\begin{array}{c}
n\\
a+b
\end{array}\right)\left(\begin{array}{c}
a+b\\
a
\end{array}\right)}{\left(\begin{array}{c}
nj\\
(a+b)j
\end{array}\right)\left(\begin{array}{c}
(a+b)j\\
aj
\end{array}\right)}.
\]
Using Stirling's approximation, the binomial coefficient can be bounded
using\vspace{-7mm}

\begin{multline*}
\\
\frac{1}{\sqrt{2\pi n\lambda(1-\lambda)}}\exp\left(nh(\lambda)-\frac{1}{12n\lambda(1-\lambda)}\right)\le\left(\begin{array}{c}
n\\
\lambda n
\end{array}\right)\\
\le\frac{1}{\sqrt{2\pi n\lambda(1-\lambda)}}\exp\left(nh(\lambda)\right),\\
\vspace{-10mm}
\end{multline*}
where $h(\cdot)$ is the entropy function in nats \cite{Gallager-1963}.
 Applying this bound to $D$ gives, after some manipulation, that\vspace{-7mm}

\begin{multline*}
\\
j\exp\Biggl((1-j)\biggl(nh\left(\frac{a+b}{n},1-\frac{a+b}{n}\right)+\\
(a+b)h\left(\frac{a}{a+b},\frac{b}{a+b}\right)\biggr)-\frac{1}{12n}\Biggr)\le D\\
\le j\exp\Biggl((1-j)\biggl(nh\left(\frac{a+b}{n},1-\frac{a+b}{n}\right)+\\
(a+b)h\left(\frac{a}{a+b},\frac{b}{a+b}\right)\biggr)+\frac{1}{12jn}\Biggr).\\
\vspace{-10mm}
\end{multline*}
Finally, we notice that\vspace{-7mm}

\begin{multline*}
\\
nh\left(\frac{a+b}{n},1-\frac{a+b}{n}\right)+(a+b)h\left(\frac{a}{a+b},\frac{b}{a+b}\right)=\\
nh\left(\frac{a}{n},\frac{b}{n},1-\frac{a}{n}-\frac{b}{n}\right).\\
\vspace{-10mm}
\end{multline*}
This completes the proof.
\end{IEEEproof}
\bibliographystyle{ieeetr}

\end{document}